\documentclass[fleqn,usenatbib]{mnras}

\usepackage{newtxtext,newtxmath}

\usepackage[T1]{fontenc}

\DeclareRobustCommand{\VAN}[3]{#2}
\let\VANthebibliography\thebibliography
\def\thebibliography{\DeclareRobustCommand{\VAN}[3]{##3}\VANthebibliography}

\usepackage{graphicx}	
\usepackage{amsmath}	
\usepackage{mathtools}
\usepackage{booktabs}
\usepackage{longtable}
\usepackage{cuted}
\usepackage{placeins}
\usepackage{enumitem}
\setlist[enumerate]{itemsep=0mm}
\usepackage{float}
\usepackage{soul}
\usepackage{hyperref}
\usepackage{orcidlink}
\usepackage{float}

\title[Striped Jets Simulations]{Jets with Streaks: the Global Dynamics and Structure of Striped Poynting flux dominated Jets}

\author[H. S. Chan et al.]{
Ho-Sang Chan\,\orcidlink{0000-0003-2776-082X}, $^{1,2}$\thanks{E-mail: hschanastrophy1997@gmail.com}
Anna Chashkina,$^{1}$
Omer Bromberg\,\orcidlink{0000-0003-4271-3941},$^{3}$
Gregory R. Werner,$^{4}$
Hayk Hakobyan\,\orcidlink{0000-0001-8939-6862},$^{5}$
\\
$^{1}$JILA, University of Colorado and National Institute of Standards and Technology, 440 UCB, Boulder, CO 80309-0440, USA\\
$^{2}$Department of Astrophysical and Planetary Sciences, University of Colorado, 391 UCB, Boulder, CO 80309, USA, \\
$^{3}$The Raymond and Beverly Sackler School of Physics and Astronomy, Tel Aviv University, Tel Aviv 69978, Israel, \\
$^{4}$Center for Integrated Plasma Studies, Physics Department, University of Colorado, 390 UCB, Boulder, CO 80309, USA, \\
$^{5}$Center for Computational Astrophysics, Flatiron Institute, 162 5th Avenue, New York, NY 10010, USA
}

\date{Accepted 2026 February 26. Received 2026 January 31; in original form 2025 November 7}

\pubyear{\the\year{}}

\begin{document}
\label{firstpage}
\pagerange{\pageref{firstpage}--\pageref{lastpage}}
\maketitle

\begin{abstract}

Striped jets, characterized by magnetically dominated regions with alternating polarity, can enable fast magnetic energy dissipation in global current sheets formed in between the stripes, and are thus considered as a viable source for the high-energy emission observed in blazars and gamma-ray burst prompt emission. We perform the first global axisymmetric GRMHD simulations that follow the evolution of striped jets from their launching, by accretion of poloidal magnetic loops with alternating polarity onto a spinning black-hole, to their energy dissipation at large distances. The accretion of each loop launches a Poynting-flux-dominated mini-jet carrying a helical magnetic field whose polarity is inherited from the loop, and the collection of which forms the jet core. The relativistic core is surrounded by a non-relativistic sheath composed of returning fieldlines from previously launched mini-jets in a nested configuration. Efficient mini-jet launching requires poloidal loops in the accretion disk to form close to the black hole ($r \lesssim 50\,r_g$); otherwise, they lose coherency due to turbulence. Once launched, the mini-jets accelerate electromagnetically, similar to steady-state jets, with the Lorentz factor increasing linearly with the jet cross-sectional radius as long as lateral causal contact with the jet axis is maintained. Current sheets form when two trailing mini-jets carrying oppositely directed magnetic fields come into contact. Early sheet formation is facilitated by the longitudinal expansion of the mini-jets, driven by internal pressure before the flow becomes super-fast-magnetosonic. We measure an energy dissipation rate of a few tens of percent of the jet Poynting power in the observer frame, within our simulation box of $10^4\,r_g$. Our results strengthen the implication that energy dissipation in current sheets of striped jets is responsible for the bright superluminal features observed in BL Lac type blazars.

\end{abstract}




\section{Introduction} \label{sec:intro}

Relativistic jets are observed in systems involving accretion onto black holes (BHs), such as active galactic nuclei \citep{1998AJ....115.1295K, 2019ARA&A..57..467B}, X-ray binaries \citep{2001MNRAS.322...31F, 2004MNRAS.355.1105F, 2006MNRAS.371.1334R, 2022Sci...378..650K}, and gamma-ray bursts \citep{1999ApJ...519L..17S, 2002ApJ...571..779P, 2014PASA...31....8G, 2015PhR...561....1K}. They emit across the electromagnetic spectrum, from radio \citep{2017A&ARv..25....4B, 2019NatAs...3..387P, 2021A&A...649L...9R} to gamma rays \citep{1997A&A...324..395B, 2003ApJ...597..186S, 2012MNRAS.421.2956Z}, offering a wealth of observational features that probe the underlying plasma and radiation physics.

Efficient jet launching often requires a large-scale, coherent poloidal magnetic field threading a rapidly spinning BH, enabling the extraction of the BH rotational energy in the form of Poynting flux \citep{1977MNRAS.179..433B, 2004ApJ...611..977M, 2006MNRAS.368.1561M}. Such a requirement faces several challenges in the astrophysical context: in collapsars, the stellar core magnetic field is dominated by toroidal field due to strong differential rotation with only weak large-scale poloidal component \citep{2002A&A...381..923S, 2005ApJ...626..350H, 2006ApJ...637..914W}. In stellar convective regions, dynamos can amplify the magnetic field to near-equipartition, but it typically produces small-scale, turbulent field with little net flux \citep{2004ApJ...614.1073B, 2016ApJ...829...92A, 2024ApJ...976L..13G}. Large-scale poloidal magnetic field may be available at large distances from the BH in AGNs or X-ray binaries, however the accretion disks in these objects are believed to be geometrically thin to distances of $10^4-10^5$ gravitational radii \citep{1992AIPC..254..146N, 2017arXiv171203949V}, so that magnetic outward diffusion dominates over inward advection \citep{1994MNRAS.267..235L, 1996ApJ...473..403H}, provided that the disk is turbulent, driven by the magnetorotational instability \citep[MRI,][]{1991ApJ...376..214B}.

Alternatively, relativistic jets may be powered by small scale magnetic loops. Such a field may be generated in stellar cores by a convective dynamo, or in the accretion disk itself through a turbulent dynamo driven by MRI and buoyancy \citep{2020MNRAS.494.3656L, 2024MNRAS.532.1522J, 2026ApJ..1007L..41C}. When plasma is accreted onto the BH, it drags the field with it due to high conductivity, attaching one part of the loop to the BH horizon while the other remains connected to the disk. Differential rotation leads to the opening of the field, allowing for the Poynting flux to flow along the (poloidal) field-lines and feed the jet \citep{2005ApJ...620..889U}. To obtain an efficient energy extraction, the loop size needs to be larger than a few tens of the BH gravitational radius \citep{2021MNRAS.508.1241C, 2024MNRAS.532.1522J}. If a second loop carrying an oppositely oriented field is advected inward, it will compress the field of the first loop, causing it to self-reconnect and make room for the fieldlines of the the second loop to become connected to the BH and launch a jet with opposite polarity \citep{2015MNRAS.446L..61P, 2020MNRAS.494.4203M, 2021MNRAS.508.1241C, 2024A&A...691A.313C}. Sequential accretion of such loops, each with opposite polarity, results in the formation of a striped jet \citep{2023ApJ...954...40K, 2026ApJ..1007L..41C}.

Beyond jet launching, the mechanisms by which jets dissipate magnetic energy are central to explaining their associated electromagnetic observables, particularly for luminous jets from blazars \citep{2008ApJ...675...71S} and gamma-ray bursts \citep{2004RvMP...76.1143P}. In Poynting‑flux–dominated (PFD) jets, the dissipation of magnetic fields requires the formation of current sheets and/or turbulence that can efficiently convert magnetic energy into thermal energy \citep{2012MNRAS.419..573M, 2015MNRAS.450..183S} and accelerate electrons into a nonthermal distribution \citep{2025ARA&A..63..127S}. Current sheets can be generated at Kelvin-Helmholtz unstable jet boundaries \citep{2021ApJ...907L..44S} and/or by the jet's kink-instability \citep{2012ApJ...757...16M, 2019ApJ...884...39B, 2020ApJ...896L..31D, 2021ApJ...908..193M, 2022MNRAS.510.2391B}. Such mechanisms can explain the existence of stationary features in jets, such as the HST-1 knot in M87 \citep{2003ApJ...586L..41H, 2006MNRAS.370..981S, 2006ApJ...640..211H}, or limb brightening in Blazar jets \citep{2009ApJ...690L..31P, 2010ApJ...723.1150P, 2014ApJ...785...53N}, but they cannot account for superluminal radio features observed at parsec scales in many BL-Lac objects \citep{2011ApJ...729...26M} or the inferred highly relativistic motion of the emission regions in gamma-ray bursts \citep{2004RvMP...76.1143P, 2015PhR...561....1K, 2019pgrb.book.....Z}. Striped jets offer a complementary way to dissipate magnetic energy in the current sheets that form at regions where the toroidal field direction flips \citep{2010ApJ...725L.234L, 2025ApJ...984...28G}. As we will show here, such energy dissipation can occur over a wide range of distances, and can account for both the superluminal features at parsec scales distances in blazars as well as the prompt GRB emission regions.

Despite growing interest in the community in the potential of striped jets to explain some observed properties of relativistic jets, our understanding of these jets remains far from complete. Theoretical models either focus on the launching process \citep{2015MNRAS.446L..61P, 2020MNRAS.494.4203M, 2021MNRAS.508.1241C, 2023ApJ...954...40K, 2024A&A...691A.313C}, or study the large scale evolution of the jets using simplified 1D analytic models \citep{2002A&A...387..714D, 2002A&A...391.1141D, 2019MNRAS.484.1378G, 2021MNRAS.502.1145Z, 2023MNRAS.523.6250D}. In fact, to date there is no self-consistent model that encompasses the evolution of striped jets consistently from launching to the dissipation regions. As a result, several fundamental questions remain unanswered, such as (i) what is the global structure of striped jets, (ii) how do they accelerate, (iii) under what conditions do regions with reversing magnetic field develop current sheets, (iv) what is the energy dissipation rate in the current sheets and how it relates to the observed power of astrophysical jets?

Here, we build upon the work of \citet{2021MNRAS.508.1241C} to study the properties of striped jets using 2D general relativistic magnetohydrodynamic (GRMHD) simulations with high effective resolution and applying an adaptive mesh refinement (AMR) with $4$ levels of refinement. The series of simulations presented in this work spans four orders of magnitude in radius, providing the first detailed study of a striped jet at large distances from the BH. The paper is organized as follows. Section~\ref{sec:method} describes the simulation methodology and initial conditions. Section~\ref{sec:result} presents the main findings. Section~\ref{sec:discuss} discusses the implications of our results. Section~\ref{sec:conclude} summarizes our conclusions.

\section{Methods} \label{sec:method}

\subsection{GRMHD Simulations} \label{subsec:grmhd}
\begin{figure*}
    \centering
    \includegraphics[width=1.0\linewidth]{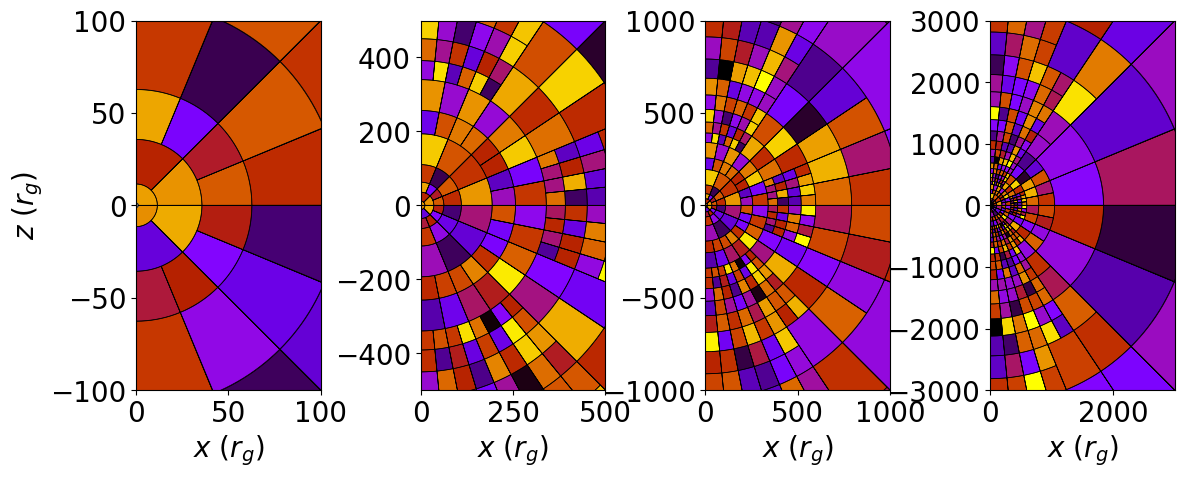}
    \caption{Mesh‑block distribution at $t = 10^4\,t_g$ for the $N_{\rm loop} = 3$ model, with all AMR levels activated. Refined blocks are concentrated along the pole, where higher resolution is required to resolve the jet at large distances from the BH. At the base resolution, each patch contains $256$ radial cells and $192$ polar cells. \label{fig:amr}}
\end{figure*}

To model the jets we use the ideal GRMHD, GPU-accelerated code  \texttt{H-AMR} \citep{2022ApJS..263...26L}, which solves the mass and energy-momentum flow equations together with the Maxwell equations in a conservative scheme (using $c=G=M_{BH}=1$): 
\begin{equation} \label{eqn:grmhd}
\begin{aligned}
    \nabla_{\mu}(\rho u^{\mu}) &= 0, \\
    \nabla_{\mu}(T^{\mu}_{\;\;\nu}) &= 0, \\
    \nabla_{\mu}(F^{*\mu\nu}) &= 0,
\end{aligned}
\end{equation}
where $T^{\mu}_{\;\;\nu}$ and $F^{*\mu\nu}$ are the stress-energy and the electromagnetic tensors defined as:
\begin{equation} \label{eqn:tensor}
\begin{aligned}
    T^{\mu\nu} &= (\rho h + b^{2})u^{\mu}u^{\nu} + (p + \frac{b^{2}}{2})g^{\mu\nu} - b^{\mu}b^{\nu}, \\
    F^{*\mu\nu} &= b^{\mu}u^{\nu} - b^{\nu}u^{\mu},
\end{aligned}
\end{equation}
with $g_{\mu\nu}$ being the metric tensor, $\rho$ being the fluid's rest mass-density and $h = 1 + \gamma p/[(\gamma - 1)\rho]$ denotes the specific enthalpy, in which $p = u_g(\gamma - 1)$ is the gas pressure, $u_g$ is the proper internal energy density and $\gamma=4/3$ is the adiabatic index. Additionally, $b^{\mu}$ is the magnetic $4-$vector, such that $b^{2} = b^{\mu}b_{\mu}$ and $u^\mu$ is the 4-velocity. In this formalism $-u_t=\sqrt{-g_{tt}}\,\Gamma$ represents the total energy per unit mass of a fluid element and approaches the Lorentz factor $\Gamma$ far from the BH.

The equations are solved in the spherical-polar form of the Kerr-Schild coordinates $(t, r, \theta, \phi)$ and  are advanced in time using a second order Runge-kutta method with a Courant-Friedrichs-Lewy number of $0.7$. The BH has a constant spin parameter of $a = +0.9375$. Length and time scales are normalized by $r_g=GM/c^2$ and $t_g=r_g/c$ respectively, while energy density quantities are normalized by $Mc^2/r_g^3$. To avoid any influence from the finite domain boundaries, we restrict our analysis to $10^4\,t_g$.

We impose constant floors on the gas density $\rho$ and internal energy density $u_{\rm g}$ such that $\rho > \max(10^{-7} r^{-2.5}, 10^{-20})$ and $u_{\rm g}  > \min(10^{-9} r^{-2.5 \gamma}, 10^{-20})$. In addition, we impose dynamic floors on $\rho$ and $u_{\rm g}$ so that $b^{2}/\rho < 100$, $b^{2}/u_{\rm g} < 1000$, $u_{\rm g}/\rho < 100$, and require that the Lorentz factor with respect to the normal observer $\Gamma_{\rm norm} < 80$. The computational domain spans the full range of polar angles under the assumption of axisymmetry. The inner radial boundary is located at $r = 1.21$\,$r_g$, covering $5$ cells inside the inner event horizon, while the outer radial boundary is set at $r = 10^4\,r_g$. The radial coordinate is constant in the logarithmic space, so that $\delta r/r$ is constant, with $\delta r$ being the radial cell size. 

The base resolution is $1024 \times 384$. We employ $4$ levels of AMR, resulting in an effective resolution of $16384 \times 6144$. The AMR criteria are designed to capture the bipolar jets at large distances and are activated only when the plasma magnetization $\sigma = b^{2}/\rho > 10^{-6}$ and the gas entropy $P/\rho^{\gamma} > 10^{-2}$. The first AMR level is activated at $r > 100\,r_g$, the second at $r > 200$\,$r_g$, the third at $r > 400$\,$r_g$, and the fourth at $r > 800$\,$r_g$. This hierarchical refinement strategy ensures that excessive grid density is avoided at small radii, where the timestep would otherwise be limited by the $r\delta\theta$ constraint. To illustrate the effect of the AMR criteria, we show the resulting mesh block distribution at the highest AMR level in Figure~\ref{fig:amr}. Using this strategy, the polar grid cells at $r = 10^4\,r_g$ can be as small as $\delta r = 5.50\,r_g$ and $r\delta\theta = 5.11\,r_g$ at the maximal level of refinement, allowing us to resolve the current sheets in the jet by $\gtrsim 5$ cells.

\subsection{Initial Conditions} \label{subsec:initial}
\begin{figure}
    \centering
    \includegraphics[width=1.0\linewidth]{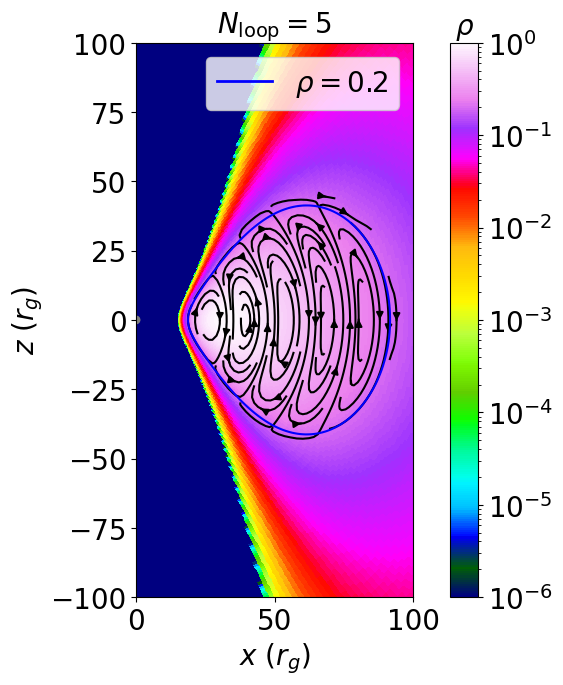}
    \caption{Initial distribution of density and magnetic field lines (black) for the $N_{\rm loop} = 5$ model. A blue contour marking $\rho = 0.2$ is also included, which closely traces the boundary of the magnetic field loops. \label{fig:initial_field}}
\end{figure}

The initial conditions are given by the Fishbone-Moncrief torus equilibrium solution \citep{1976ApJ...207..962F}. The torus is defined by an inner radius $r_{\rm in}$ and a radius of maximum pressure $r_{\rm max}$, both measured from the coordinate origin along the midplane. We adopt $r_{\rm in} = 15$\,$r_g$ and $r_{\rm out} = 32$\,$r_g$ \citep{2021MNRAS.508.1241C}. A magnetic field with multiple alternating loop topology is superimposed on the equilibrium torus, described by a vector potential of the form\footnote{See also \citet{2020MNRAS.495.1549N, 2023MNRAS.522.2307J, 2024A&A...688A..82J, 2026arXiv260825763K} for a different implementation of the multiple-loop setup.}
\begin{equation}
\begin{aligned}
  A^{\phi} &\propto (\rho - \rho_{\rm cut})\sin(\pi r_{\rm norm}N_{\rm loop})\cos(\theta), \\
  r_{\rm norm} &= \frac{r/r_{\rm mag,in} - 1}{r_{\rm mag,out}/r_{\rm mag,in} - 1},
\end{aligned}
\end{equation}
where $N_{\rm loop} = 2$, $3$, or $5$ denotes the number of loops and $\rho_{\rm cut} = 0.2$ is the cut-off density below which no magnetic field exists. In addition, $r_{\rm mag,in} = 18$\,$r_g$ and $r_{\rm mag,out} = 92$\,$r_g$ denote the inner and outer radii of the magnetized region of the torus, respectively. These boundaries are defined by the condition $\rho_{\rm cut} \geq 0.2$, and are measured along the midplane. The magnetic field is normalized so that the maximal magnetic field satisfies $b^2_{\rm max}/2=p_{\rm max}$. Such strong magnetization is chosen to accelerate the onset of accretion onto the BH, thereby enabling the formation of multiple striped jets within the simulation duration. As such, the dynamics may not represent a long-lived quasi-steady disk. The initial density distribution and magnetic-field configuration are shown in Figure~\ref{fig:initial_field}.

\section{Results} \label{sec:result}

\subsection{Accretion Dynamics and Energy Extraction} \label{subsec:tseries}
\begin{figure*}
    \centering
    \includegraphics[width=1.0\linewidth]{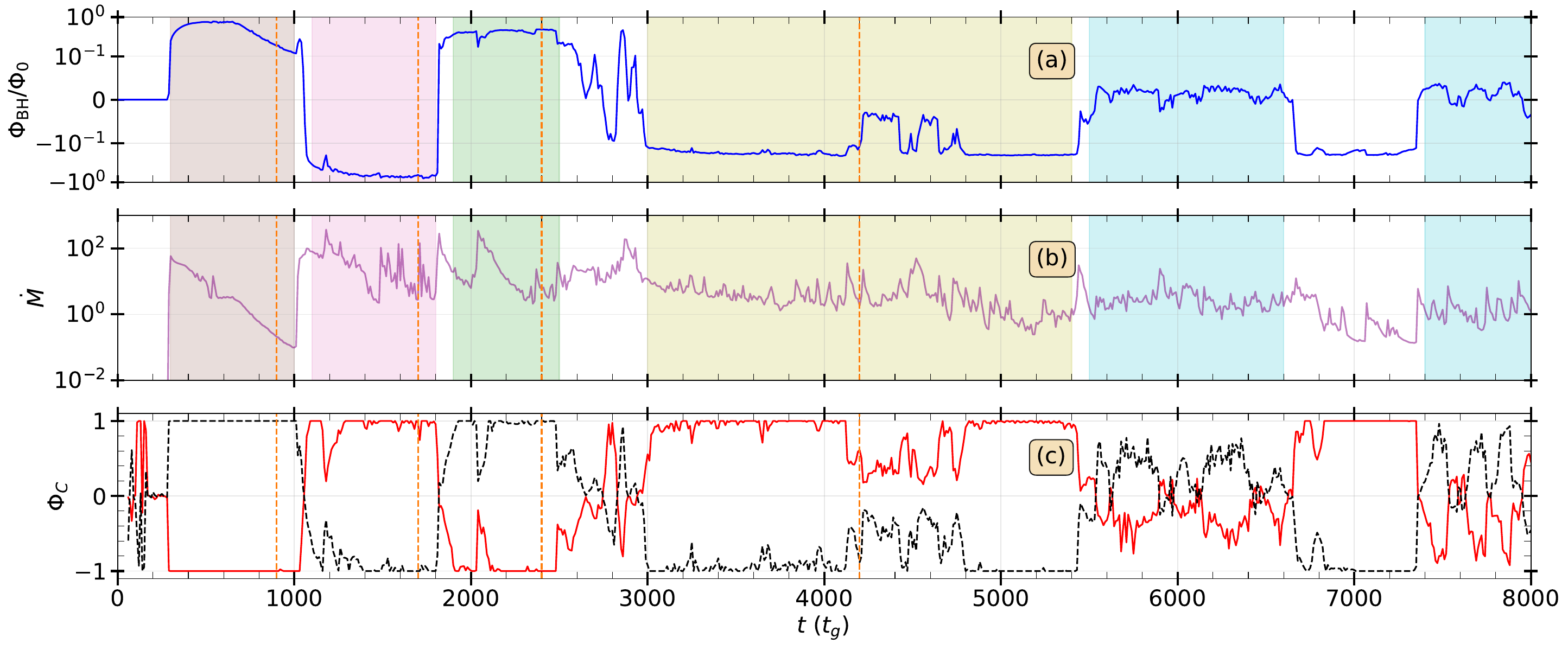}
    \caption{Time series for the $N_{\mathrm{loop}} = 5$ model. Panel~(a) shows the normalized signed magnetic flux threading the event horizon. Panel~(b) shows the mass accretion rate measured at the event horizon (purple). Panel~(c) shows the magnetic field coherency parameter for the upper (black) and lower (red) hemispheres of the BH horizon. The background is shaded to indicate periods during which a magnetic loop threads the BH: brown (first loop), pink (second loop), green (third loop), yellow (fourth loop), and blue (fifth loop). The unshaded region between $\sim 6600 - 7400\,t_g$ corresponds to the accretion of fossil loops that survived previous reconnection. \label{fig:tseries2}}
\end{figure*}
\begin{figure*}
    \centering
    \includegraphics[width=1.0\linewidth]{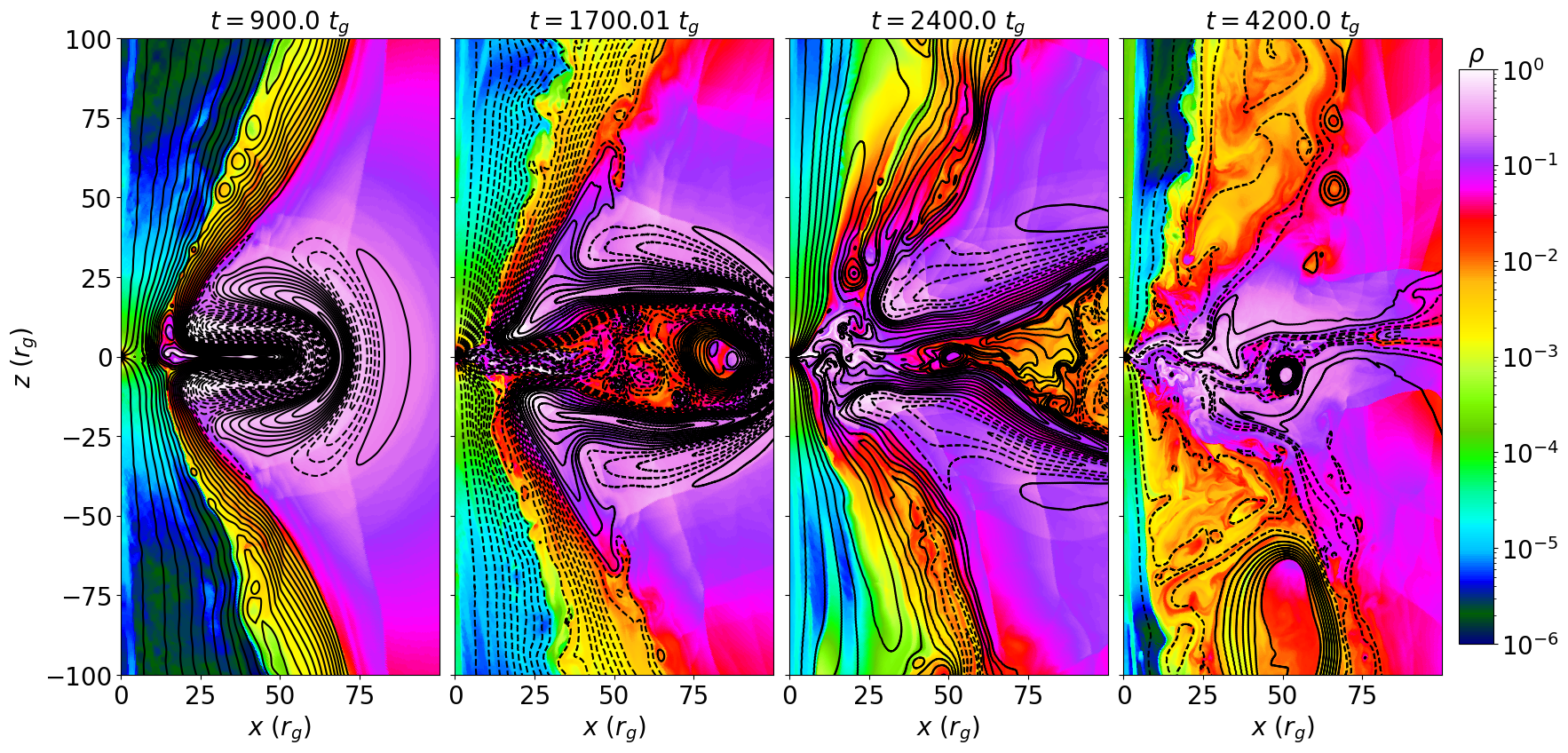}
    \caption{Density snapshot for the $N_{\rm loop} = 5$ model, with magnetic field lines (black) appended. The title of each subplot indicates the time at which the snapshot is taken, and roughly coincides with the moment when each of the first four loops is accreted onto the BH. We distinguish magnetic-field loops with opposite polarity by using different linestyles: solid and dashed. \label{fig:rho_contour}}
\end{figure*}
\begin{figure*}
    \centering
    \includegraphics[width=1.0\linewidth]{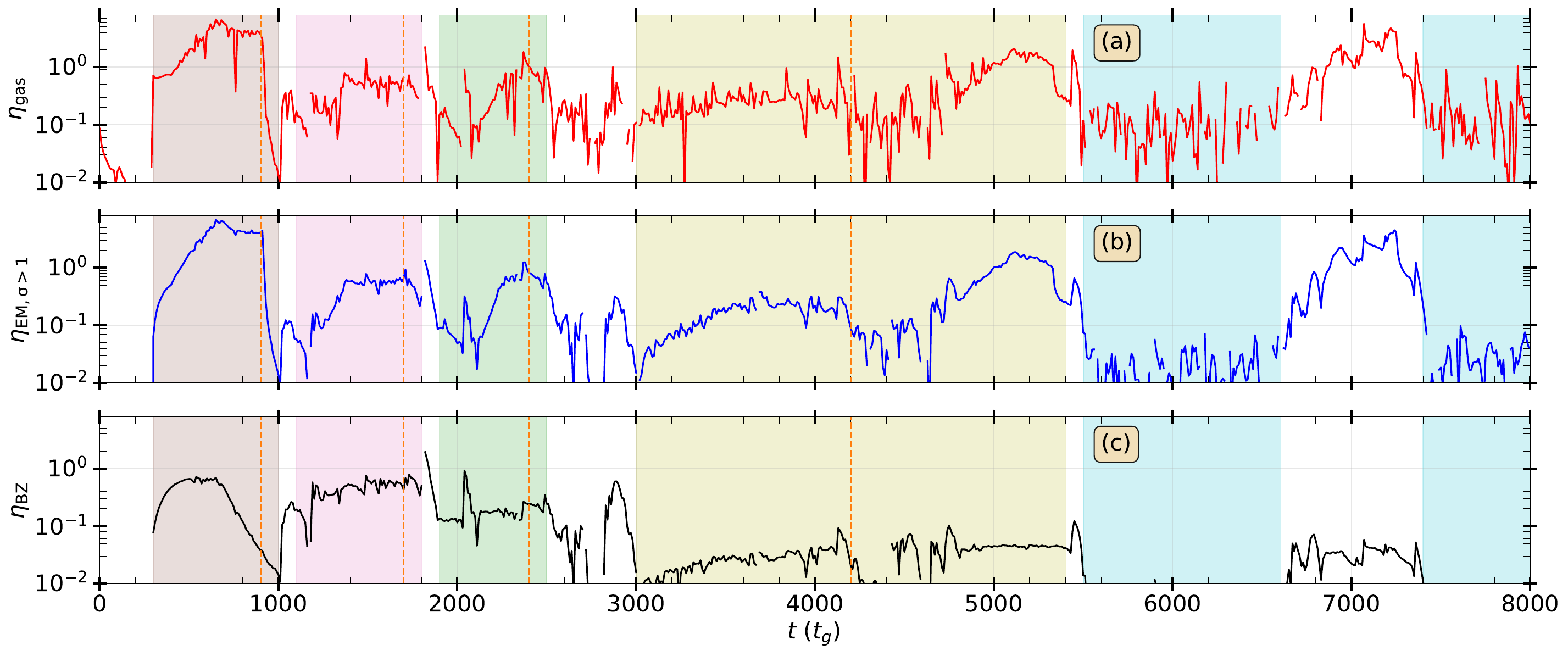}
    \caption{Same as Figure~\ref{fig:tseries2}, but for the following quantities: Panel~(a) shows the horizon gas efficiency, Panel~(b) shows the electromagnetic energy efficiency, while Panel~(c) shows efficiency with respect to the maximum BZ power. \label{fig:tseries22}}
\end{figure*}

We start by analyzing the accretion dynamics and its interplay with the buildup of fluxes on the BH horizon. We define the fluxes 
of mass ($\dot{M}$) and (signed) magnetic field ($\Phi_{\rm BH}$) as \citep{narayan2012grmhd}:
\begin{equation}
\begin{aligned} 
    \dot{M} &= -\int_{0}^{2\pi}\int_{0}^{\pi}\rho u^{r}\sqrt{-g}d\theta d\phi,
\end{aligned}
\end{equation}
and
\begin{equation}
\begin{aligned} 
    \Phi_{\rm BH} &= \left\{ \int_{0}^{2\pi}\int_{0}^{\pi/2}B^{r}\sqrt{-g}d\theta d\phi,\right. \\
                  & - \left.\int_{0}^{2\pi}\int_{\pi/2}^{\pi}B^{r}\sqrt{-g}d\theta d\phi \right\},
\end{aligned}
\end{equation}
respectively, where $B^{i} = F^{*i0}$ is the magnetic field in the frame of a normal observer and $\sqrt{-g}$ is the Jacobian determinant used to convert the equations from curvilinear coordinates to flat space, in which $g=|g_{\mu\nu}|$. The fluxes are evaluated at $r_{\rm BH} = 1 + \sqrt{1 - a^{2}}$. 

Figure~\ref{fig:tseries2} shows the time evolution of the fluxes in the $N_{\rm loop} = 5$ model. We highlight the time interval during which each loop threads the BH with a distinct color. Note that while we focus our analysis on the $N_{\mathrm{loop}} = 5$ model, we find similar behaviour in the time series for all models. Panel~(a) shows the magnetic flux expressed in units of the initial magnetic flux of the torus' strongest loop (the second loop in Figure~\ref{fig:initial_field}) 
\begin{equation}  
    \Phi_{0} = 2\int_{r_{l,1}}^{r_{l,c}} \int|B^{\theta}|\sqrt{-g}drd\phi.
\end{equation}
Here, $r_{l,1}$ and $r_{l,c}$ denote the inner radius and the center of the magnetic loop, respectively and the factor 2 is added to account for the unsigned flux on both horizon hemispheres. Panel~(b) shows the mass accretion rate $\dot{M}$, in code units, and panel (c) shows the coherency of the magnetic field threading the horizon at the northern (red) and southern (black) hemisphere, defined as \citep{2024ApJ...960...97R, santhiya2025dynamojetinterconnectionsgrmhd}
\begin{equation} 
  \Phi_{c, \theta < \pi/2} = \frac{\int\int_{r = r_{\rm BH}, \theta < \pi/2} B^r \sqrt{-g}d\theta d\phi}{\int\int_{r = r_{\rm BH}, \theta < \pi/2} |B^r| \sqrt{-g}d\theta d\phi}.
\end{equation}
Absolute values close to $1$ imply high coherency.

During the accretion of the first three loops (colored brown, pink, and green), the magnetic flux and magnetic-field coherency remain high. The mass accretion rate is highly variable, dropping by $2 - 3$ orders of magnitude while the magnetic flux on the BH retains the same sign. The gradual drop in $\dot{M}$ during the accretion of the first loop is simulative, caused by the formation of a magnetic bubble in the disk midplane during the onset of the simulation, which pushes plasma away from the BH as it moves outward under buoyancy forces (see Figure~\ref{fig:rho_contour}). 

The variability in $\dot{M}$ during the accretion of the second and third loops is more typical to a magnetically arrested disk state \citep[MAD;][]{2003PASJ...55L..69N,2008ApJ...677..317I,2011MNRAS.418L..79T,2012MNRAS.423.3083M}, associated with the build-up and saturation of magnetic flux threading the BH. Such a build-up of flux can exert a significant magnetic pressure that overcomes the gas ram pressure and impedes mass inflow. When an opposite polarity loop is advected inward, it compresses the magnetic field of the currently accreted loop, squeezing it and forcing it to self-reconnect. This temporarily removes the magnetic barrier and allows for previously stalled material to accrete. This state is visualized by sharp rises in $\dot{M}$ seen at $t \sim 1000$ and $1800\,t_g$, associated with a flip in the signed magnetic flux. At later times ($t \gtrsim 2500\,t_g$), during the accretion of the fourth and fifth loops, the magnetic field in the disk becomes turbulent and dominated by small-scale structure, with significantly reduced overall coherence. This in turn reduces the accumulation of magnetic flux on the BH, weakening the magnetic barrier and allows for more gradual mass accretion. The accretion flow transitions into the standard and normal evolution state \citep[SANE;][]{narayan2012grmhd}, while the aforementioned variability largely vanishes.

The process by which polarity switches on the BH horizon and the development of turbulent magnetic fields is illustrated in Figure~\ref{fig:rho_contour}, which shows four snapshots of the inner disk at times marked in the magnetic-flux evolution plots of Figure~\ref{fig:tseries2} by vertical orange lines. At $t = 900\,t_g$, the innermost loop is attached to the BH through its outgoing field lines (green-blue region) and to the inner disk through its returning field lines (yellow-red region). A current sheet lies between them, identified by a chain of plasmoids. When the second loop, marked by dashed lines in the inner pink region, is advected inward, it cannot reconnect with the first loop because at the contact surface of both loops carry magnetic fields of the same direction. However, the second loop exerts magnetic pressure and compresses the first loop, forcing it to self-reconnect. The magnetic flux on the horizon then drops, making room for the second loop to attach to the BH and re-establish jet launching with magnetic fields of flipped polarity ($t = 1700\,t_g$).

Figure~\ref{fig:rho_contour} also demonstrates equatorial current sheets being ejected away from the BH. The sheets are produced by self‑reconnection of the accreting loop. As discussed above, the initial transient magnetic bubble pushes plasma away from the BH, causing the magnetic loop to flip against itself and deform ($t = 900\,t_g$). The process also creates a low‑density void near the midplane. As a result, the subsequent loop cannot be accreted along the midplane and instead travels inward along off-midplane directions. When these distorted loops merge near the BH, the upper half of the loop carries magnetic field lines of opposite direction to those in the lower half, promoting self‑reconnection and altering the field topology. The self‑reconnection then ejects a current sheet along the midplane, which in turn further pushes and deforms the subsequent loops ($t = 1700\,t_g$), creating a vicious cycle. The heat injected into the plasma also drives turbulence and further twists the initially coherent poloidal field ($t = 2400\,t_g$–$4200\,t_g$). The loss of magnetic-field coherency renders jet launching inefficient. Consequently, the jet opening angle at the jet base progressively decreases between $t = 900\,t_g$ to $t = 4200\,t_g$. The outward transport of the loop's center also explains why the fourth loop threading the BH survives for an extended period (see Figure~\ref{fig:tseries2}), as the fifth loop requires additional time to be advected from large radii to the BH.

Last, we look at the energy extraction efficiency. We define the total energy flow $\dot{E}$ and the EM energy flow, $\dot{E}_{\rm EM}$ as:
\begin{equation}
\begin{aligned} 
    \dot{E} &= \int_{0}^{2\pi}\int_{0}^{\pi}T^{r}_{t}\sqrt{-g}d\theta d\phi, \\
    \dot{E}_{\rm EM} &= \int_{0}^{2\pi}\int_{0}^{\pi}T^{r}_{t, \rm EM}\sqrt{-g}d\theta d\phi, 
\end{aligned}
\end{equation}
where $T^{r}_{t}$ is the total radial energy flux term of the stress-energy tensor (Equation~\ref{eqn:tensor}) and $T^{r}_{t,{\rm EM}} = T^{r}_{t} - \rho h u^{r} u_{t}$ is the radial EM energy flux. All quantities are evaluated at $r_{\rm BH}$. Based on these fluxes we define three types of efficiencies \citep[see also][]{2011MNRAS.418L..79T}:
\begin{equation}
\begin{aligned} 
    \eta_{\rm gas} &= (\dot{M} - \dot{E})/ \langle\dot{M}\rangle, \\
    \eta_{_{\rm{EM},\sigma>1}} &= \dot{E}_{\rm EM, \sigma >1}/\langle\dot{M}\rangle,\\ 
    \eta_{_{\rm BZ}} &= \dot{E}_{\rm EM, \sigma >1}/\dot{E}_{\rm BZ}, 
\end{aligned}
\end{equation}
where $(\dot{M} - \dot{E})$ represents the total energy flow minus the rest-mass energy flow, and $\dot{E}_{\rm EM, \sigma >1}$ is the electromagnetic energy flow carried by strongly magnetized matter with $\sigma > 1$. Both quantities are normalized by the mass accretion rate averaged over a time window of $200\,t_g$. The quantity $\eta_{_{\rm BZ}}$ represents the electromagnetic power output normalized by the expected BZ power obtained from the magnetic field of the strongest loop in the disk, defined as $\dot{E}_{\rm BZ} = \Omega_{\rm BH}^2 \Phi_{0}^2/(24\pi^2 c)$, where $\Omega_{\rm BH} = ac/(2r_{\rm BH})$ is the angular frequency of the BH.

We show $\eta_{\rm gas}$, $\eta_{_{\rm{EM},\sigma>1}}$, and $\eta_{_{\rm BZ}}$ in Figure~\ref{fig:tseries22} (a), (b), and (c), respectively. These efficiencies exhibit a similar pattern. The first three loops are highly efficient at converting accretion power into outflow and extracting the BH energy. In particular, the first loop attains both outflow and electromagnetic efficiencies well above unity, extracting energy at a rate close to the theoretical maximum predicted by the BZ process ($\eta_{\rm BZ} \sim 1$). The second and third loops have lower efficiencies than the first loop, yet they maintain efficiencies of $\lesssim 1$.

The outflow efficiency drops during the accretion of the fourth and fifth loops, consistent with the decline in magnetic-field coherency and the reduced absolute amount of magnetic flux threading the BH. In particular, both the outflow and electromagnetic efficiencies fluctuate between positive values and negative (accretion-dominated) values, while the BZ efficiency falls below $0.1$ --- there is insufficient net poloidal flux to extract energy from the BH and power outflows. Although in some cases, episodes of large-scale field accumulation on the BH horizon lead to temporarily high values of $\eta_{\rm gas}$ and $\eta_{\rm EM,\sigma>1}$, and $\Phi_{c} \sim 1$, for example at times $4800\,t_g < t < 5400\,t_g$, the absolute magnetic flux on the horizon remains comparatively small, and the jet is therefore underpowered ($\eta_{\rm BZ} \ll 1$). Still, the jet power is of order $\mathcal{O}(10)$\,\%, meaning that a magnetic field with flipping polarity can convert at least $10$\,\% of the accreted mass energy into outflows.

\subsection{Striped-jet Properties} \label{subsec:properties}
\begin{figure*}
    \centering
    \includegraphics[width=1.0\linewidth]{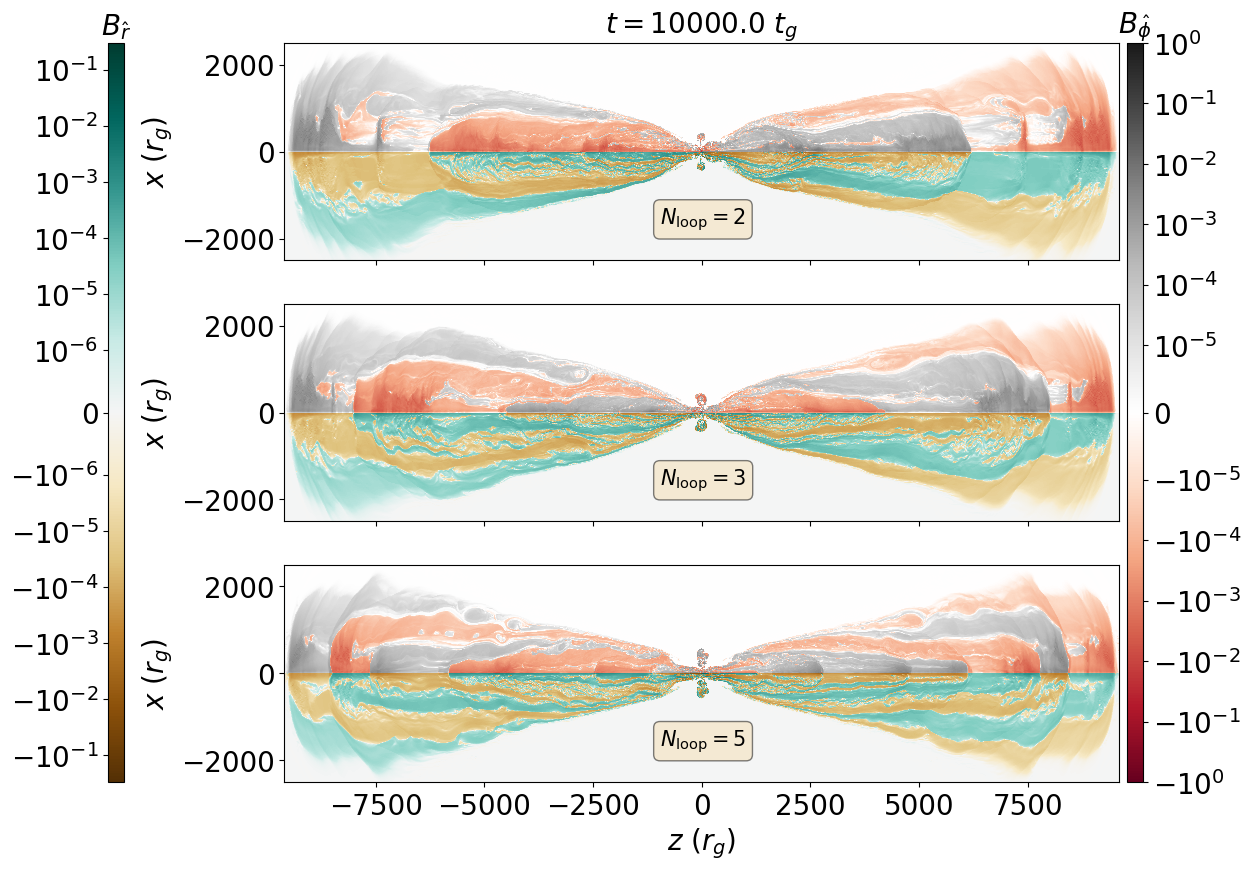}
    \caption{Maps of the toroidal (red-black) and radial (green-gold) magnetic fields (in the quasi-orthonormal frame) along the $x-z$ plane. The first, second, and third rows show the results for the $N_{\rm loop}=2$, $3$, and $5$ models, respectively (see textbox). The snapshot is taken at $t = 10^4\,t_g$. 
    \label{fig:bphi}}
\end{figure*}
\begin{figure*}
    \centering    
    \includegraphics[width=1.0\linewidth]{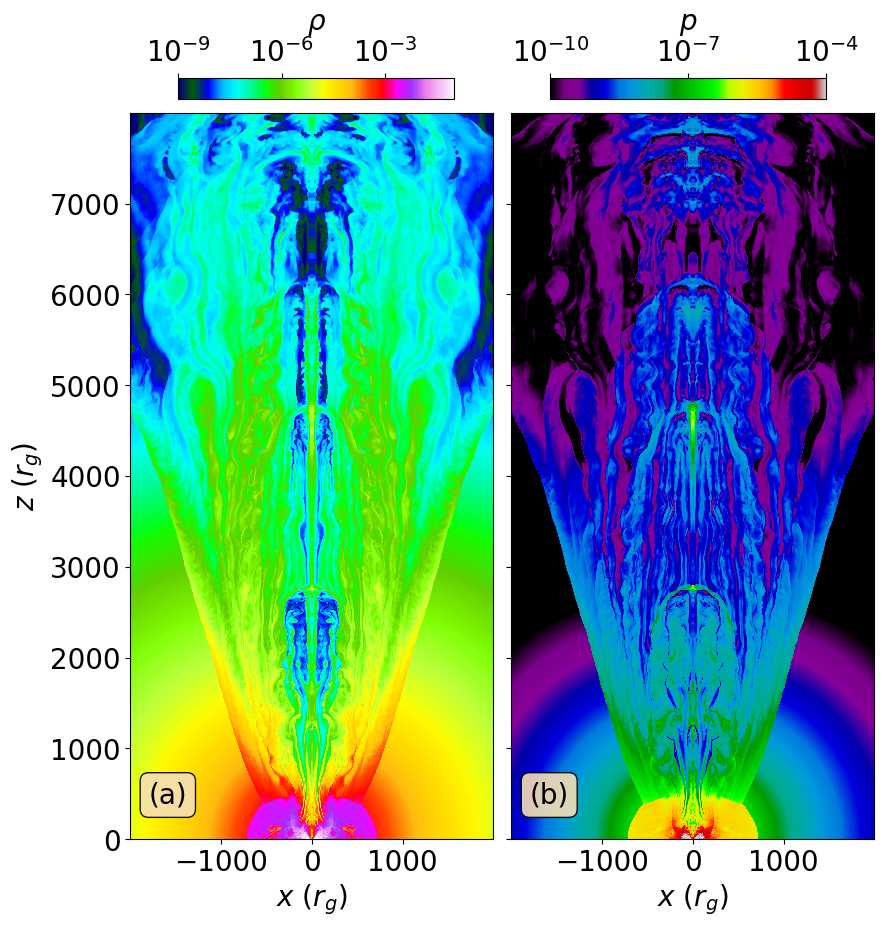}
    \caption{Maps of (a) the density and (b) the gas pressure at $t = 10^4\,t_g$ in the $x-z$ plane for the $N_{\rm loop}=5$ model. The maps are mirrored across the $x=0$ line, and only the upper hemisphere is shown for simplicity. \label{fig:pgas}}
\end{figure*}
\begin{figure*}
    \centering
    \includegraphics[width=1.0\linewidth]{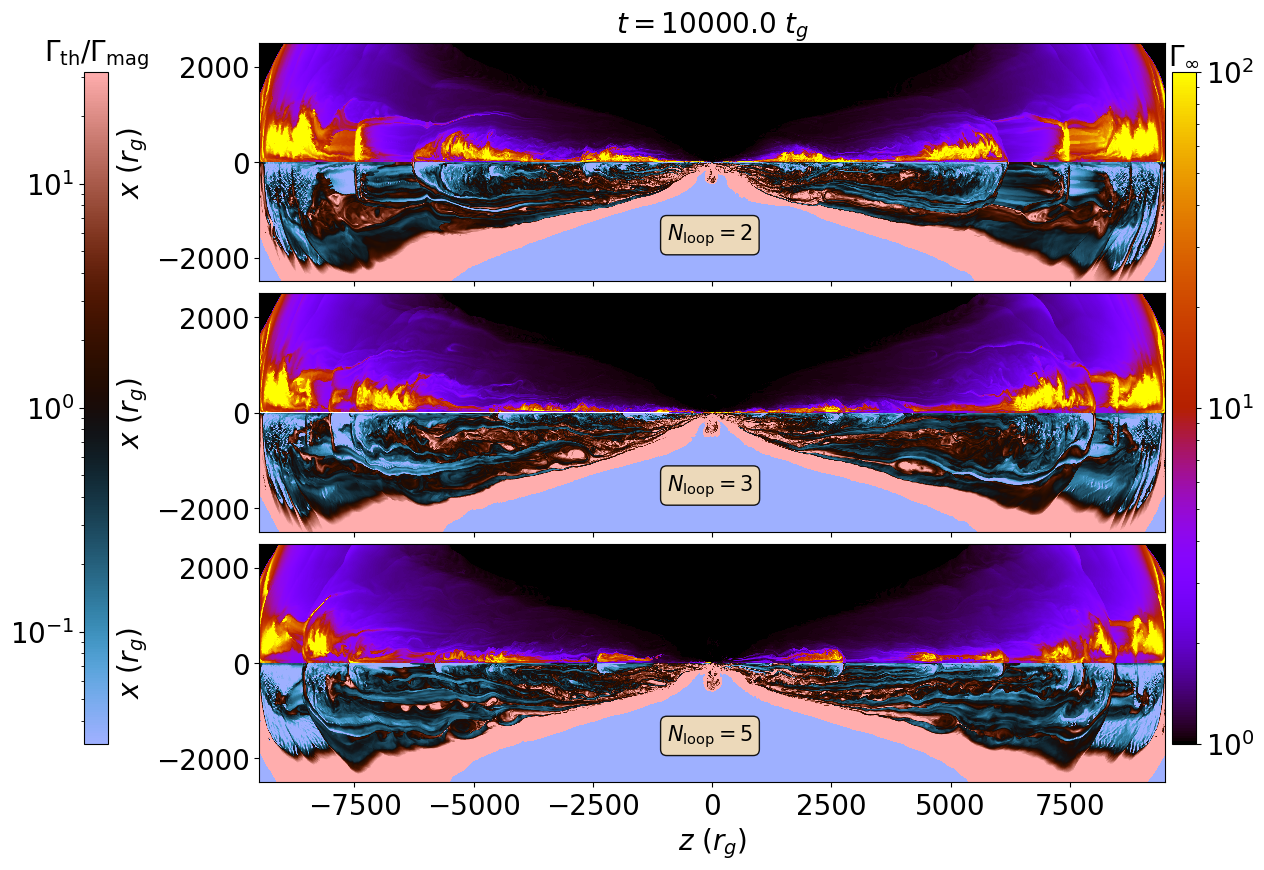}
    \caption{Same as Figure~\ref{fig:bphi}, but showing the total asymptotic Lorentz factor ($\Gamma_{\infty}$) and the ratio between thermal and magnetic energy density equivalent to $(h-1)/\sigma$. The mini-jets along the spine have low baryon content and are characterized by high values of $\Gamma_{\infty}$, while the sheath has a higher baryon load and shows filaments of low thermal to magnetic energy density ratio that tracks the returning fieldlines from each jet launching episode.} \label{fig:gamma_infty}
\end{figure*}
\begin{figure*}
    \centering
    \includegraphics[width=1.0\linewidth]{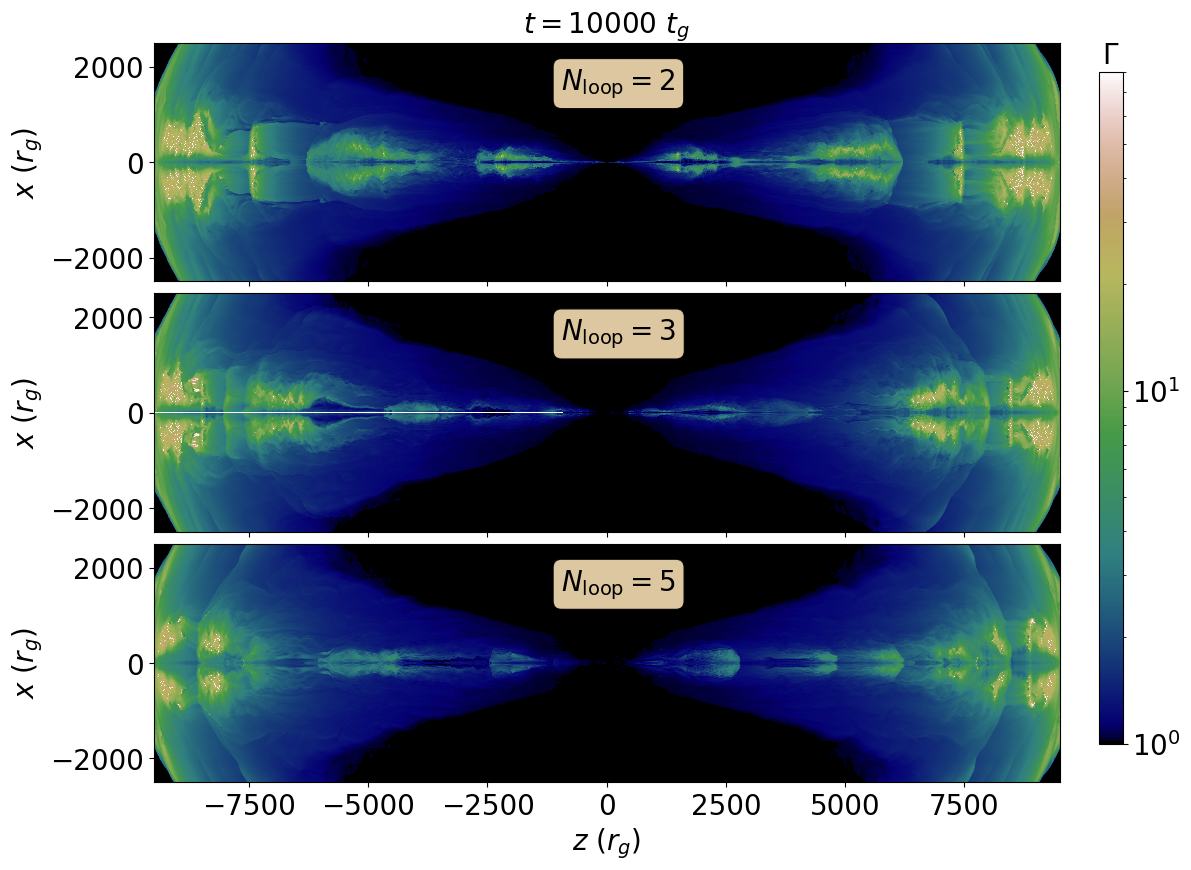}
    \caption{Same as Figure~\ref{fig:bphi}, but showing the gas Lorentz factor. 
    \label{fig:additional}}
\end{figure*}

During the consecutive loop accretion, each jet-launching episode sends a `mini-jet' that propagates into the relics of the previous jet, resulting in a `jet-within-a-jet' configuration. Figure~\ref{fig:bphi} demonstrates this configuration by showing the spatial distribution of the radial and toroidal field in our three models at $t = 10^4\,t_g$, when the tip of the first jet reaches the edge of the computational box. The jet spine consists of the field lines corresponding to the leading edge of a loop, which carry the electromagnetic power from the BH. It is composed of $\sim N_{\mathrm{loop}}$ mini-jets with oppositely oriented fields in models initialized with $N_{\mathrm{loop}}$ poloidal loops. The jet sheath contains the returning field lines of loops from previous launching episodes in a nested configuration. Analytic works of striped jets \citep{2002A&A...387..714D, 2019MNRAS.484.1378G, 2023MNRAS.523.6250D} typically assumed a prescribed longitudinal width for the stripes and did not account for the horizontal structure. Because the jet expands and accelerates, anticipating the full multidimensional structure of the stripes is not-trivial. In addition, the nested jet configuration enforces a lateral structure, which was not accounted by such models.

We identify cases where the stripes are in tight contact with each other producing thin current sheets. They are characterized by sharp reversals in the toroidal-field sign over length scales far smaller than the mini-jet width, like in the case of the first $3$ mini-jets launched during the MAD state of the $N_{\rm loop}=5$ model. In other cases, as in the $N_{\rm loop}=2$ model, the stripes are separated by regions of weak magnetic field whose spatial extent is comparable to the mini-jet length. Such separation arises from the time lag between the end of the previous and the start of the new jet launching phase. In particular, the separation between the third and fourth mini-jets in the $N_{\rm loop} = 5$ model is due to the fact that the fourth and fifth loops break into smaller scale structures by turbulence and thus it takes longer time for sufficient flux to be built on the BH horizon for an efficient jet launching (e.g., $t = 2500 - 3000\,t_g$ in Figure~\ref{fig:tseries2}). The conditions leading to the formation of stripes in tight contact are discussed in later sections.

Figure~\ref{fig:pgas} shows the rest-mass density (left) and pressure (right) of the plasma in the $N_{\rm loop} = 5$ model. The jet is launched into a conical funnel that forms along the axis due to the non-spherical geometry of the accretion disk and extends up to $z \leq 6000\,r_g$. The jet spine appears as the low-density region along the axis and is composed of several mini-jets trailing one another. The mini-jet launched by the second loop at $z > 7000\,r_g$ has cleared out of the funnel and is surrounded by a tenuous, cold medium. Below are the mini-jets launched by the third ($5000\,r_g \leq z \leq 6000\,r_g$), fourth ($1500\,r_g \leq z \leq 5000\,r_g$), and fifth ($z \leq 1000\,r_g$) loops. They propagate in a denser medium filled by matter from earlier launching episodes. The propagation leads to the formation of bow shocks that compress the returning fieldlines of the mini-jets into narrow cocoons \citep{2020MNRAS.498.3320G, 2021MNRAS.504.3947G}, seen as pockets of hot, high-density plasma around the lower mini-jets. The cocoons maintain a pressure balance with the mini-jets. As the pressure profile in the cocoons is rather uniform  the mini-jets that propagate in the funnel assumes a cylindrical morphology. Such configuration is similar to that of a steady jet propagating in a stationary medium \citep{2011ApJ...740..100B, Bromberg2016}. 

The nested configuration of the jet sheath is visible as vertical filaments of high-pressure and high density material inside the funnel. We identify roughly four such filaments encircling the jet core. The origin of the baryons is matter that slips in during episodes of polarity switch on the BH, when the drop in magnetic pressure allows for a sharp increase in the mass accretion rate. When magnetic flux is rebuilt on the BH and a new mini-jet emerges, it pushes the mass away from the pole and shocks it, forming a new filament. Once the mini-jets and their cocoons exit the disk-confined region at $z \gtrsim 6000\,r_g$, they expand sideways and open up.

To analyze the magnetic content of the jet we plot in Figure~\ref{fig:gamma_infty} the generalized Bernoulli constant, defined as
\begin{equation} \label{eqn:ginfty}
    \Gamma_\infty =(h + \sigma)u_t,
\end{equation}
which measures the terminal Lorentz factor of a fluid element if it expands adiabatically, and the ratio between thermal and magnetic energy density (measured as $(h-1)/\sigma$). The mini-jets along the jet spine are characterized by high magnetization and low baryon content, giving them  potential to accelerate to high Lorentz factor. Between the mini-jets there are regions of low $\Gamma_\infty$, a consequence of high baryon loading that occurs when matter streams into the BH polar region during periods of magnetic-polarity switch. It can also be seen that the peak in $\Gamma_\infty$ in each mini-jet occurs off axis, in agreement with the fact that the jet power is proportional to the strength of the toroidal field, which is small on the jet axis (see Appendix~\ref{sec:supp}). The energy density in the sheath is dominated by baryons and has a relatively low magnetic field content. These features are typical to steady‑state, Poynting-flux-dominated (PFD) jets \citep{2009ApJ...698.1570L, 2010NewA...15..749T, 2016MNRAS.456.1739B}. Here, we show that they also hold for striped jets.

Figure~\ref{fig:additional} shows the jet Lorentz factor for all models at $t = 10^4\,t_g$. In each simulation the first mini-jets launched during MAD states exhibit the highest Lorentz factors, reaching orders of a few tens. Since the terminal Lorentz factor in these mini-jets is on the order of a few hundred, it implies that most of the mini-jet energy is still carried by the magnetic field and is available for dissipation. In regions where the magnetic field changes sign, we see sharp jumps in the Lorentz factor of the flow. 

\subsection{Time-evolution and Jet Dynamics} \label{subsec:jet_accelerate}
\begin{figure*}
    \centering
    \includegraphics[width=1.0\linewidth]{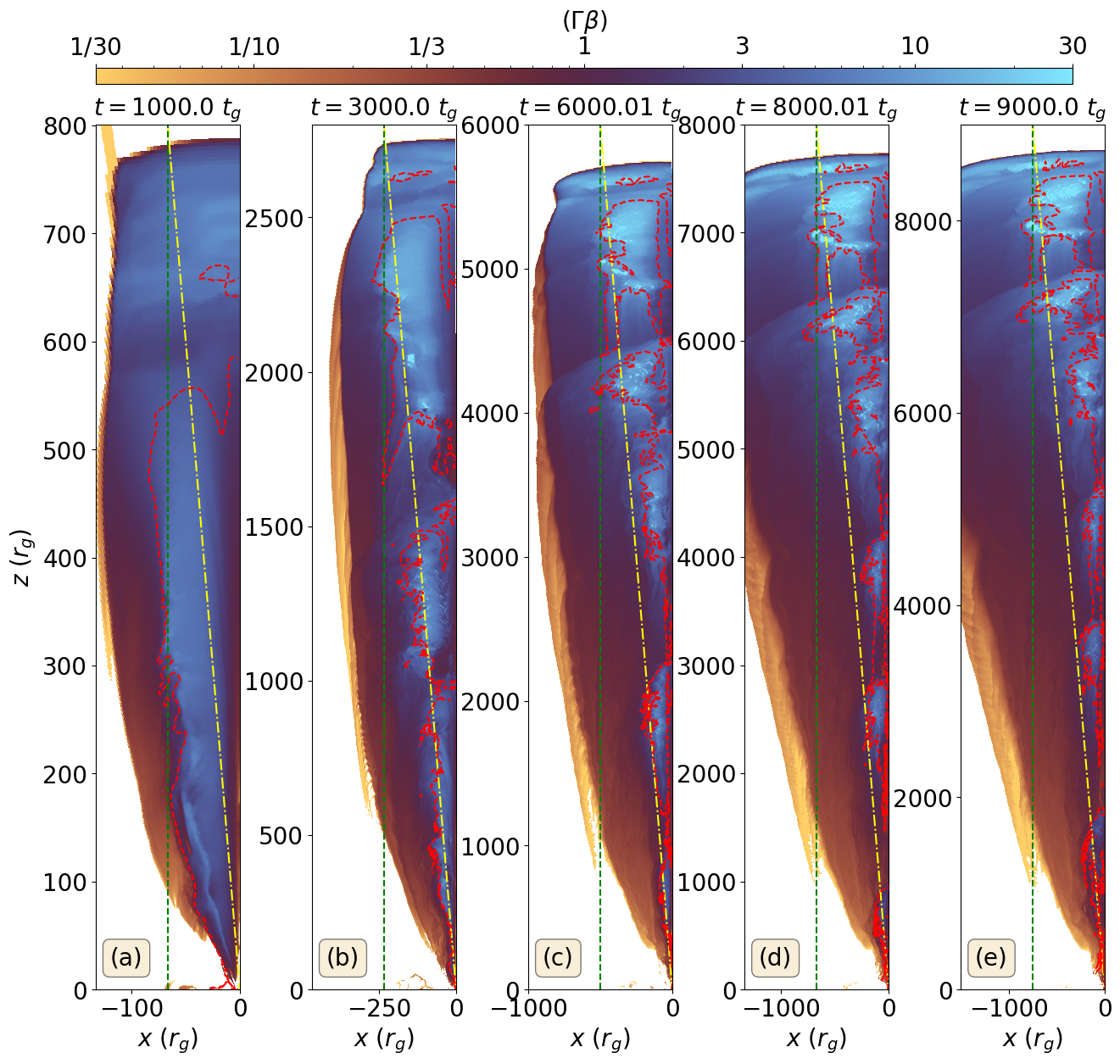}
    \caption{The proper velocity map along the $x - z$ plane at selected times. The red contour marks the $\sigma = 1$ boundary, which we define as the jet. All panels are shown with a constant aspect ratio of $6{:}1$. The vertical dashed green line is placed at the midpoint of the horizontal extent. The yellow dashed-dotted line tracks a constant opening angle of $1/12$ rad.}
    \label{fig:jetradevo}
\end{figure*}
\begin{figure*}
    \centering
    \includegraphics[width=1.0\linewidth]{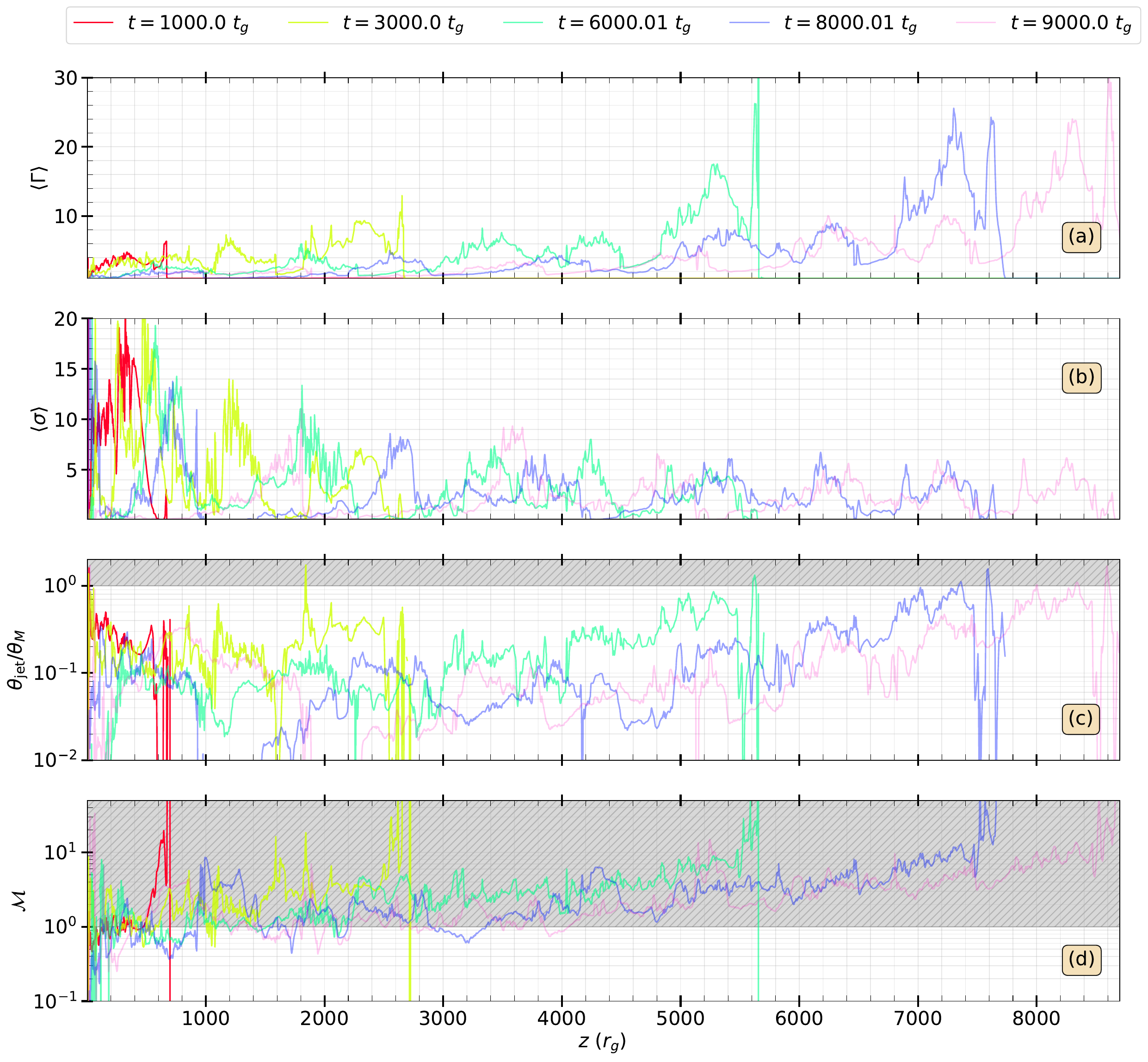}
    \caption{Panel (a): the jet-average Lorentz factor; (b): the jet-average magnetization; (c): the ratio of the jet opening angle to the Mach cone angle ; (d): the magnetic mach number for the $N_{\rm loop} = 5$ model at different times in the upper hemisphere. For (c) and (d), the shaded regions indicate where the ratio exceeds unity. Since the jet-spine opening angle is $\lesssim 1/12 \approx 0.083 \ll 1$, we have $r = z\cos\theta \approx z$. We therefore label the $x$-axis as $z$ to a good approximation. \label{fig:acceleration_profile}}
\end{figure*}
\begin{figure}
    \centering
    \includegraphics[width=1.0\linewidth]{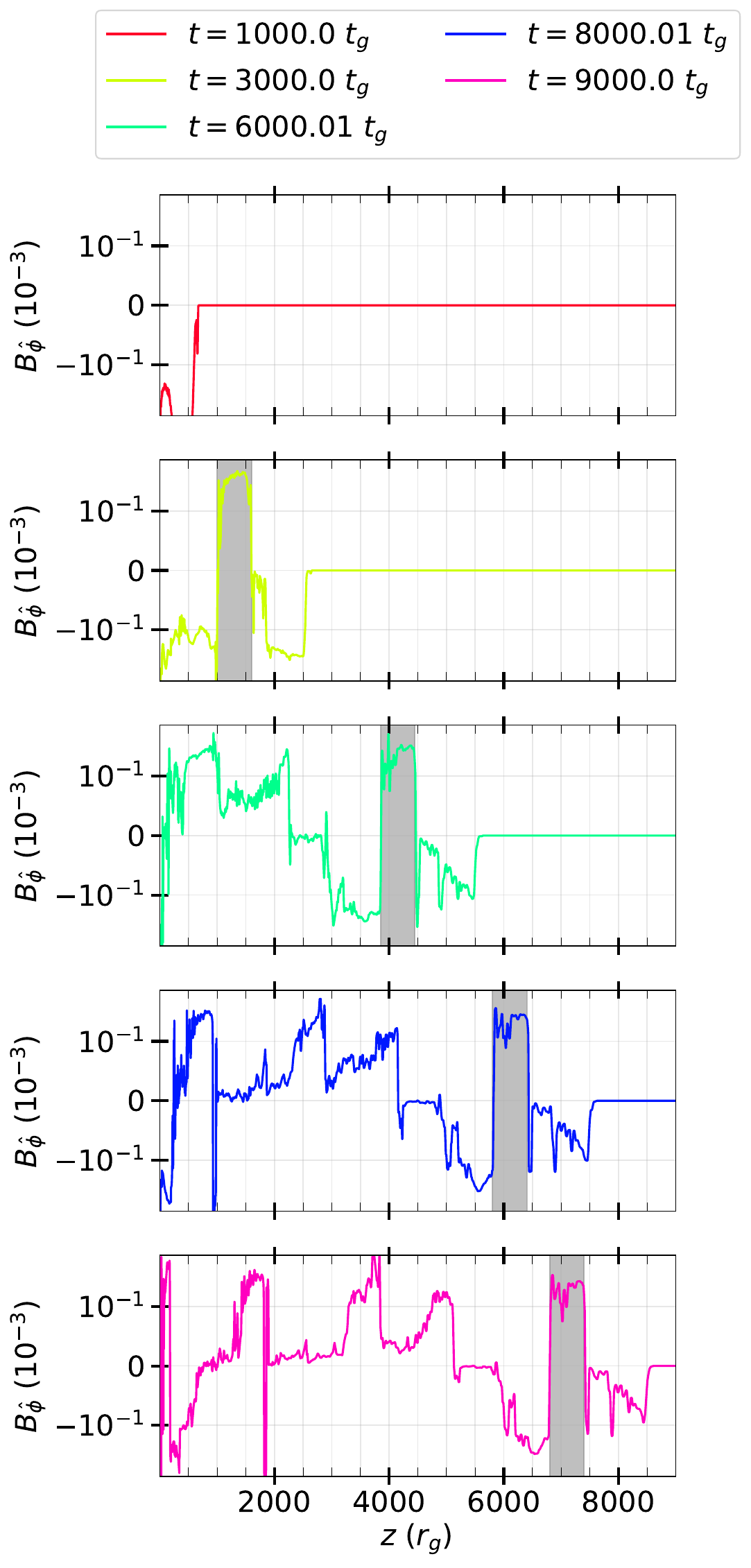}
    \caption{Same as Figure~\ref{fig:acceleration_profile}, but for the toroidal magnetic field (measured along the pole). A dark-grey rectangular box of width $\sim 600\,r_g$ highlights that the width of the second mini-jet remains roughly constant with time. Note that along this axis, $z \approx r$. \label{fig:jet_radius}}
\end{figure}

In Figure~\ref{fig:jetradevo} we show snapshots of the proper velocity for the $N_{\rm loop}=5$ case at different times. The vertical extent of each panel is $\sim c t$, while the aspect ratio in all panels is the same. Red contours outline the $\sigma\approx 1$ region, which identifies the interior of the jet. The first two mini-jets have parabolic profiles close to the launching point, as shown in panel (a), transitioning to a conical one beyond $z \simeq 1000\,r_g$, highlighted with the yellow line (panels [b]-[e]). Mini-jets launched subsequently at later times are more strongly confined and maintain roughly a constant cylindrical radius (see panels [d]-[e] at $z\lesssim 4000\,r_g$. We account this difference to the environment in which either of these mini-jets is propagating: while the first two mini-jets propagate through the low density initial disk material, later mini-jets have to force their way through the pre-contaminated axial funnel, which confines the flow in the process.

In PFD jets,  morphology is tightly connected with jet acceleration \citep{2007MNRAS.380...51K, Lyubarsky2009, 2009MNRAS.394.1182K, 2010NewA...15..749T}. Efficient acceleration requires strong causal connection with the jet axis, namely, the transverse velocity of the jet material should be smaller than the fast magnetosonic velocity. In this case the Lorentz factor is expected to be proportional to the cylindrical radius. Thus, for the first two mini-jets, their Lorentz factor should grow linearly with $z$, while for the third mini-jet its acceleration is expected to stall when jet's sideways expansion stops. Figure~\ref{fig:acceleration_profile} (a) depicts the jet's average Lorentz factor as a function of $z$ at five different times, defined as \citep{2024ApJ...964...79L}
\begin{equation} 
    \langle \Gamma \rangle = \frac{\int\int f_{\rm KE} \sqrt{-g}d\theta d\phi}{\int\int f_{\rm RM} \sqrt{-g}d\theta d\phi},
\end{equation}
where the kinetic energy flux $f_{\rm KE}$ and rest-mass flux $f_{\rm RM}$ are
\begin{equation}\label{eq:Gam_avv}
\begin{aligned} 
    f_{\rm KE} &= -\rho u^{r}u_{t}, \\
    f_{\rm RM} &= \rho u^{r}.
\end{aligned} 
\end{equation}
The integral is evaluated at the upper hemisphere and is restricted to regions with $\sigma > 1$\footnote{We use a logistic function $f(x) = 1/(1 + \exp\!\left[-\frac{x - 1}{0.05}\right])$ to smoothly transition between magnetized and non-magnetized material.} associated with the relativistic jet spine. Each mini-jet is associated with a localized bump in the Lorentz factor profile. We find that the peak Lorentz factor of the first mini-jet grows linearly with $z$ from $\Gamma \approx 5$ at $t=1000\,t_g$ to $\Gamma \approx 20$ at $t=8000\,t_g$, consistent with a conical expansion, while between $t=8000-9000\,t_g$ the acceleration stops. As we will show below the velocity freeze out is the result of a loss of causal contact with the axis. Note that the  spike in the Lorentz‑factor profile at the head of the jet is consistent with the location of the bow shock and is likely numerical.

Jet acceleration converts internal energy into bulk motion. In PFD jets it implies that $\sigma\propto\Gamma^{-1}$ ($\sigma\gg h$ in Equation~\ref{eqn:ginfty}). We plot the jet's average magnetization in Figure~\ref{fig:acceleration_profile} (b), defined by
\begin{equation} 
\begin{aligned} 
    \langle \sigma \rangle &= \frac{\int\int f_{\rm EM} \sqrt{-g}d\theta d\phi}{\int\int f_{\rm KE} \sqrt{-g}d\theta d\phi}, 
\end{aligned} 
\end{equation}
where the electromagnetic energy flux $f_{\rm EM}$ is
\begin{equation} 
\begin{aligned} 
    f_{\rm EM} &= -(b^{2}u^{r}u_{t} - b^{r}b_{t}).
\end{aligned} 
\end{equation}
the integrals are evaluated at the same regions as in Equation~\ref{eq:Gam_avv}. The first mini-jet has a peak magnetization of $\sigma=20$ at $t=1000\,t_g$, which drops to $\sigma=5$ at $t=8000\,t_g$, consistent with the increase in $\Gamma$ during these times. Between $t=8000\,t_g$ and $t=9000\,t_g$ the magnetization is unchanged. 

Last, we address the importance of causal connectivity to the evolution of the mini-jets. We define the fast magnetosonic mach number as
\begin{equation} 
\begin{aligned} 
    \mathcal{M} = \frac{\langle \Gamma \beta\rangle} {\sqrt{\langle \sigma \rangle}},
\end{aligned} 
\end{equation}
where $\langle \Gamma \beta\rangle=\sqrt{\langle \Gamma\rangle^2-1}$. The Mach angle is defined as $\Theta_{\rm M}=\mathcal{M}^{-1}$, while the jet opening angle is defined as \citep{2024ApJ...963L..29R}
\begin{equation} 
\begin{aligned} 
    \theta_{\rm jet} &= \text{arccos}\left(1 - \frac{1}{2\pi r^{2}}\int\int\sqrt{-g}d\theta d\phi\right).
\end{aligned} 
\end{equation}
In Figure~\ref{fig:acceleration_profile} (c) we plot the ratio of $\theta_{\rm jet}/\Theta_\mathcal{M}$. Values larger than $1$ are associated with super fast expansion in the lateral direction and a loss of causal connection with the axis. As it can be seen the first mini-jet becomes laterally super fast above $z = 7000\,r_g$ consistent with the freeze-out in the Lorentz factor and plasma magnetization evolution, thereby confirming that the acceleration mechanism is indeed magnetic.

Apart from the lateral direction, the mini-jets also expand in the longitudinal direction under their internal pressure. This behavior differs from that of a steady, continuous jet and modifies their acceleration profile \citep{2011MNRAS.411.1323G}. The expansion pushes the front and back ends of the mini-jet in opposite directions, increasing the acceleration of the front while reducing the acceleration of the rear. This can allow trailing mini-jets to catch up and dissipate their magnetic energy via reconnection, if they carry oppositely oriented magnetic fields. Once the flow becomes super-fast-magnetosonic, information can no longer travel backwards and pressure-driven longitudinal expansion becomes ineffective. In the absence of an external force that decelerates the flow, the mini-jets may not be able to gain on each other fast enough to interact at distances where they are still PFD and can generate effective current sheets\footnote{The mini-jets may eventually come into contact due to small differences in velocity and Lorentz factor between their leading and trailing ends. For $\Delta\Gamma\ll\Gamma$, this will take place at distances of the order $\Gamma^3/\Delta\Gamma$ times the initial separation between the mini-jets, which are much larger than our computational domain}. This limits the initial separation between successive mini-jets and thus constrains the timescale for polarity reversals on the BH and, in turn,  the size of the accreted poloidal loops. The actual value  depends on the physical properties of the system and is beyond the scope of the current study. 

Figure ~\ref{fig:acceleration_profile} (d)  shows the average Mach number in the jet at the same times as before.  As it can be seen, the flow becomes super-fast-magnetosonic at $z\gtrsim3000\,r_g$ (marked by a yellow line). To measure the longitudinal width of the mini-jets we show in Figure~\ref{fig:jet_radius} the toroidal magnetic field along the pole. Each mini‑jet can be identified as a region of coherent $B_{\hat{\phi}}$ polarity, while the boundary between mini‑jets appears as a sharp jump in $B_{\hat{\phi}}$. We mark in gray rectangle the position of the second mini-jet in each snapshot is marked with a gray rectangle, which maintains a constant width of $\Delta z=600\,r_g$ from $t=3000\,t_g$ and forth, in agreement with the super-fast-magnetosonic motion of the flow. 

\subsection{Current Sheet and Energy Dissipation} \label{subsec:sheet}
\begin{figure*}
    \centering
    \gridline{
    \fig{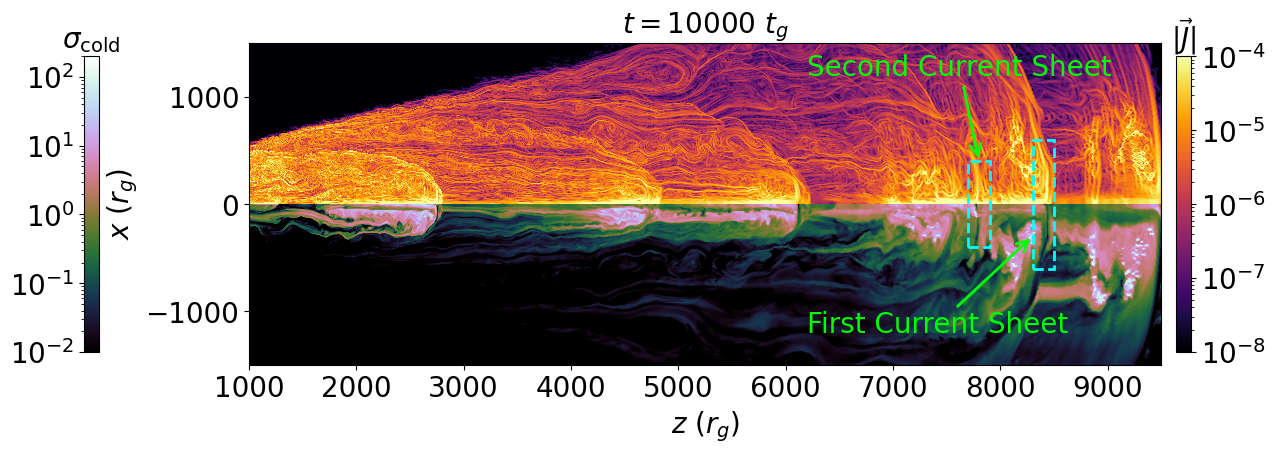}{1.0\textwidth}{(a)}}
    \gridline{
    \fig{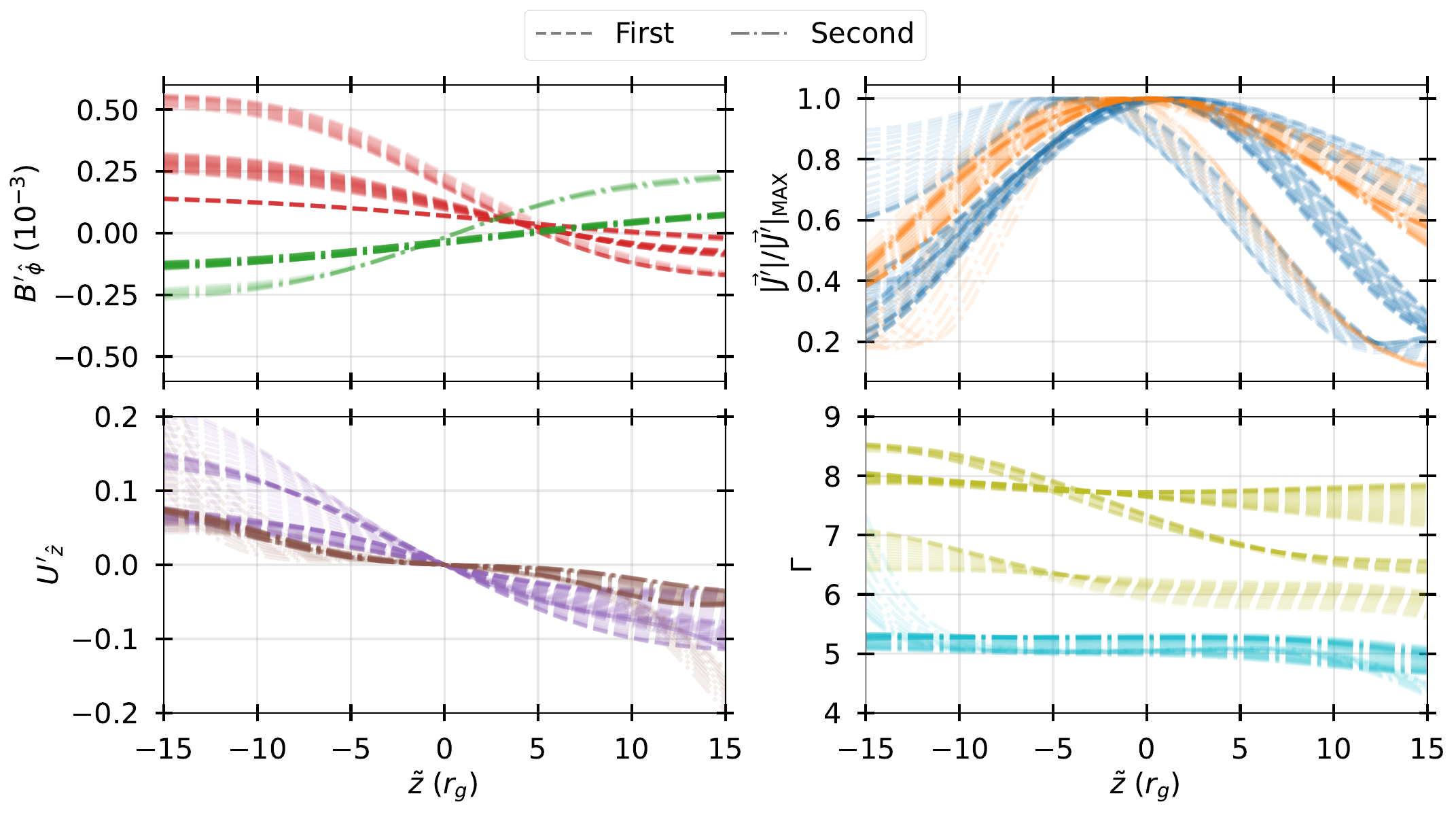}{1.0\textwidth}{(b)}}
    \caption{(a) The map shows, in the upper half, the $3-$current density and, in the lower half, the plasma magnetization in the lab frame for the $N_{\rm loop} = 5$ model at $t = 10^4\,t_g$. A rectangular box and annotation highlight the current sheet that will be examined in detail. (b) The toroidal magnetic field ($B'_{\phi}$), current density ($|\vec{J}'|$), and vertical velocity ($U'_{z}$) in the fluid frame, together with the Lorentz factor ($\Gamma$) in the lab frame, for the two current sheets we analyze (see panel [a]). The horizontal axis shows distances in the local frame of each current sheet, where $J'_{max}$ is set at $\tilde{z}=0$. The current density is normalized by its respective maximum value. \label{fig:current_sheet_local}}
\end{figure*}
\begin{figure*}
    \centering
    \includegraphics[width=1.0\linewidth]{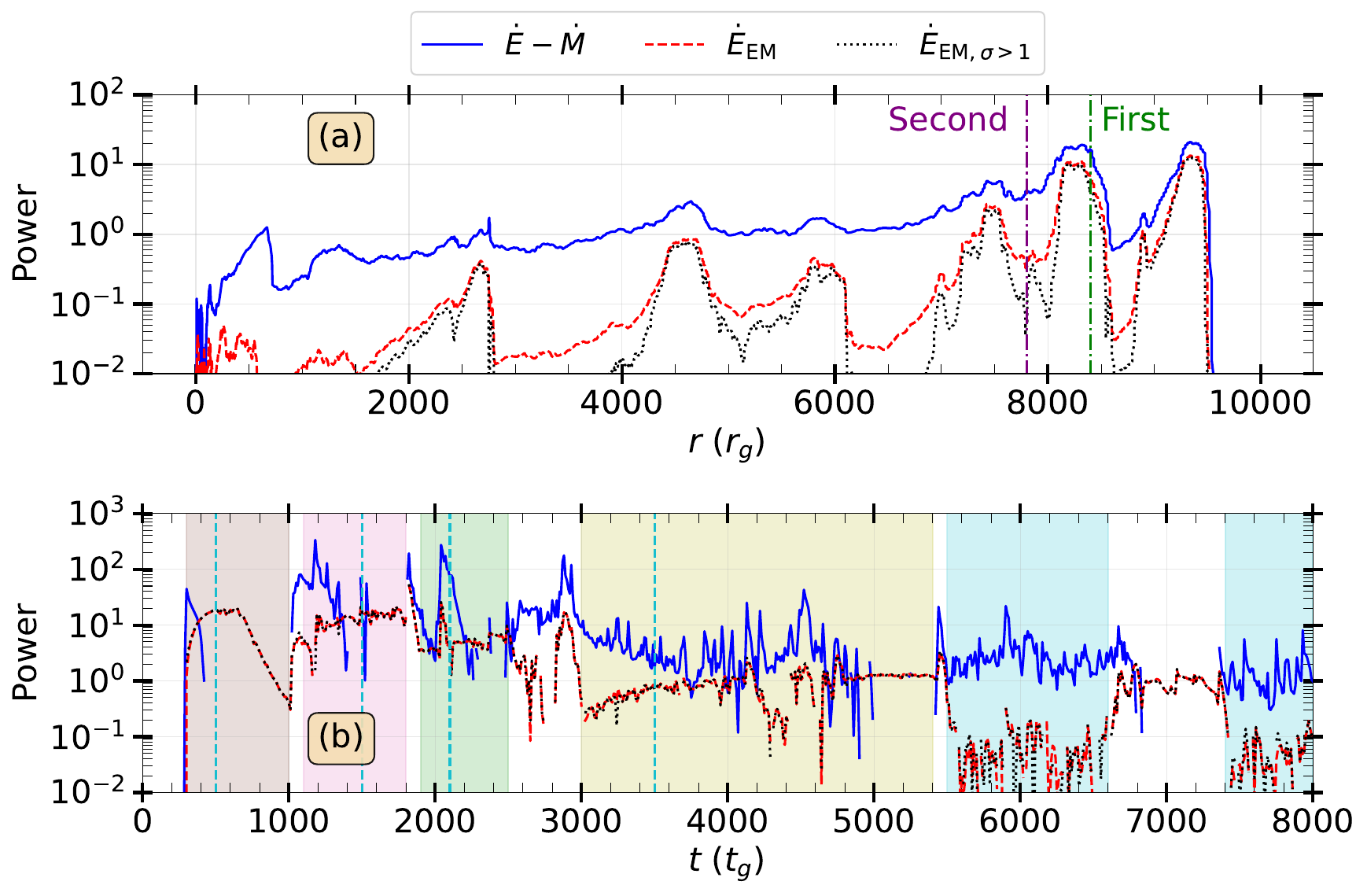}
    \caption{(a) Net power (blue; excluding rest-mass energy), electromagnetic power (red), and electromagnetic power integrated across the jet (black) emerging from the BH event horizon as a function of distance from the BH at $t = 10^4$\,$t_g$ for the $N_{\rm loop} = 5$ model. The locations of the first (green) and second (purple) current sheets (see Figure~\ref{fig:current_sheet_local} [a]) are marked with vertical lines. (b) Same as (a), but showing the power as a function of time for the upper hemisphere only. The vertical cyan line marks the approximate instant at which a mini-jet is launched, i.e., when a clean funnel forms near the BH event horizon. No such clean funnel is observed for the fifth mini-jet. \label{fig:jetpower}}
\end{figure*}

At the transition region between two trailing mini-jets with oppositely oriented field, a current layer is formed. If the mini-jets are in close contact, the current layer steepens into a current sheet. In Figure~\ref{fig:current_sheet_local} (a), we illustrate the current density (top) and magnetization (bottom) of the $N_{\rm loop} = 5$ model at time $t=10^4\,t_g$. Current sheets are seen as sharp peaks in the current density accompanied by a sharp drop in the magnetization. The two cyan dashed rectangles highlight the current sheets at the front and back of the second mini-jet launched during a MAD state, hereafter referred to as 'first' and 'second' respectively. Two other current sheets are seen at $z \simeq 6000\,r_g$ and $z = 2700\,r_g$, however they are formed around weaker toroidal field and thus are less prominent. In panel (b) we show the variation of the toroidal magnetic field, current density and proper velocity across the first (dashed) and second (dash-dotted) current sheets, measured in the frame of the sheets, together with the lab frame Lorentz factor. In Appendix~\ref{sec:fluid} we describe the method of transforming the plasma properties from the lab to the sheet frame. All values are measured along vertical trajectories, where each line in the plot represents a trajectory that samples the sheet at a different horizontal location. It can be seen that the peak of the current density is shifted from the zero point of the toroidal field. The reason for that is an existence of an additional component of magnetic field in the $\hat{R}$ direction that modifies the current distribution. 

The energy dissipation rate in the current sheets is controlled by collective non-ideal plasma processes (e.g., plasma resistivity), which determine the thickness of the diffusion region and the rate at which magnetic flux reconnects. In ideal MHD simulations, reconnection is mediated by numerical resistivity. Consequently, dissipation occurs on the grid scale and is resolution dependent. Our simulations thus provide only a gross estimate of the magnetic energy dissipation rate, and hence of the energy budget available to power the observed emission from these regions. To estimate the dissipation rate in a current sheet in our simulation we write the expression:
\begin{equation} \label{eqn:dissipation}
    \frac{dE'}{d\tau}\sim u_B\frac{dV'}{d\tau}\simeq \pi R_{_{\rm CS}}^2 B'^2\Delta v_{z},
\end{equation}
where $\tau$ is the proper time, $u_{B}$ is the magnetic energy density, $V'$ is the volume of plasma entering the current sheet from both sides, $R_{_{\rm CS}}$ is the cylindrical radius of the current sheet, $\Delta v_z=v_z^{up}+v_z^{down}$ is the sum of inflow velocities from both sides of the current sheet and we use a Heaviside normalization for the magnetic field as in the simulation. We assume that the current sheet is planar on the $x-y$ plane and all values are evaluated in the frame of the current sheet. Taking typical values for the plasma properties from Figure~\ref{fig:current_sheet_local} (b), $B'_\phi \sim 5\times10^{-4}$, $\Delta v_{z} \sim 0.2(0.1)$, and $R_{_{\rm CS}} \sim 600(400)\,r_{g}$, for the first(second) current sheet, we get a dissipation rate $\dot{E}'\sim6 \times 10^{-2} (10^{-2})$ in code units. This quantity is measured in the frame of the current sheet. To convert it to an observed power we need to transform it to the observer frame. Taking a typical Lorentz factor of $\Gamma\sim 8(5)$ we get an observed jet power $L_j = \eta\Gamma^2\dot{E}' \sim 3.6(0.3)\eta$ for the first(second) current sheet, where $\eta\leq1$ is the fraction of dissipated energy that is converted to light (see Appendix~\ref{sec:CS_Power} for a proper derivation). 

We can compare the observed power with the jet Poynting power, calculated in two ways, as shown in Figure~\ref{fig:jetpower}. In panel (a) we calculate the power as a function of $r$ at a time $t = 10^4\,t_g$, while in panel (b) we calculate the power on the BH horizon as a function of time. The times corresponding to the launch of each mini-jet are marked with shaded colors, and the orange dashed vertical lines correspond to the instances when the head and tail of the second mini-jet, which later evolve into the current sheets, were launched. In both cases it can be seen that the EM power of the first and second mini-jets is of the order $10$ in code units while that of the third mini-jet is of the order $3$, making the EM jet power a factor $(3-10)/\eta$ larger than our estimated $L_j$. This implies that the current sheets dissipate enough energy to power the bright superluminal features observed in blazar jets, or  even the prompt GRB emission, if it is created by such mechanism.

\section{Discussion} \label{sec:discuss}

\subsection{Implications} \label{subsec:imply}

The dissipation process in our model continues as long as fresh magnetic field is supplied to the current sheet from both sides. We can evaluate the proper lifetime of the current sheet by taking the longitudinal scale of a mini-jet in the frame of the current sheet and dividing it by the proper inflow velocity into the current sheet. Multiplying that by the Lorentz factor of the current sheet, we get an estimate for the duration over which the current sheet can be active in the BH frame
\begin{equation}
    \Delta t_{_{\rm CS}}\simeq \Delta l'_{_{\rm mj}} \Gamma_{_{\rm CS}}/v'_{z}.
\end{equation}
The length scale of the mini-jet is determined by the accretion time of the poloidal loop used to launch it, which in our $N_{\rm loop} = 5$ model is of order $\Delta l'_{_{\rm mj}}\sim 10^3\,r_g$, assuming the jet travels with a bulk speed of $v/c \sim 1$.  Taking a typical inflow velocity of $v'_{z}\sim0.1c$ and a Lorentz factor $\Gamma_{_{\rm CS}}\sim10$ we get $\Delta t_{_{\rm CS}}\sim10^5\,t_g$. Furthermore, since $B'_\phi\propto R^{-1}$ and the energy dissipation rate scales like $\dot{E}_{_{\rm CS}}\sim {B'}^2_\phi R_{_{\rm CS}}^2v'_{z}\Gamma^2$, it is expected to be roughly constant if the mini-jet is coasting. In the case of blazar jets it implies that we expect to find superluminal features over large distances, ranging from close vicinity to the BH up to $10^4-10^5\,r_g$, equivalent to $1-10\,\text{pc}$ for a $10^{9}M_\odot$ BH.

A possible example for a system hosting such a striped jet may be the TeV blazar TXS 0506+056 \citep{Paiano2018}. Radio observations of the system show an extended jet structure reaching up to $\sim7$ mas ($\sim35$ pc) from the core, which features $4$ radio components where the furthest component (\#1 at $4$ mas) and the closest component (\#4 at $\sim 0.5$ mas) to the core move superluminaly with an apparent velocity $\lesssim2c$ \citep{Homan2021,Eppel2026}. Component \#4 first appeared at 2019, about $2$ years after a major TeV flare event at the core region that was associated with a VHE neutrino detection \citep{IceCube2018}. In addition, long-term VLBA observations of the jet revealed a coherent wave of enhanced radio emission propagating along the inner $\sim2$ mas ($\lesssim10$ pc) part of the jet with an apparent velocity of $21c$ \citep{Kovalev2026}.

One interpretation of this observation is that the jet is composed of a highly relativistic spine surrounded by a mildly relativistic sheath. The observed radio emission is dominated by the sheath, which contains the mildly relativistic features. Interaction of the spine with the sheath generates a dissipative pattern that moves with the spine and generates the excess coherent fast radio component \citep{Kovalev2026}. If the spine is formed by a striped jet, the multimessenger TeV event may signal the launching of a mini-jet, while the onset of the radio emission at a $2$ years delay can mark the crossing of the radio photosphere. In this picture the fast superluminal wave pattern may form by matter that enters the current sheet with the reconnecting magnetic field at the head of the mini-jet, and gets evacuated along the radial flux tubes and interacts with the surrounding sheath. The observed distance of the wave from the core of $1-10$ pc fits our estimation for the lifetime of the current sheet.

\subsection{3D vs 2D} \label{subsec:3D}

In reality, non-axisymmetric 3D effects will modify the evolution of the current sheets. 3D reconnection of toroidal field generates a radial current that breaks into radial filaments via tearing instability (3D version of plasmoids). Differential rotation in the jet bends the filaments in the toroidal direction, imprinting an Archimedean spiral geometry on them \citep{Lyubarsky2001, Petri2016}. Consequently, secondary 3D instabilities, such as current-driven kink instability, are expected to disrupt the filaments, driving the current layer into a fully turbulent state \citep{Zenitani2007, Sironi2014, Werner2021, Berta2025, French2026, Deka2026}. This scenario resembles a relativistic extension of the reconnection layer in striped pulsar wind, but with a crucial distinction: the turbulent current sheet is continuously energized by a fresh influx of magnetized material entering from above and below the current layer. The turbulent current layer creates a stochastic structure of local reconnection sites moving in random directions within the fluid frame instead of a global sheet. The resultant emission will be composed of multiple, locally beamed and randomly oriented emission sites that can account for the fast, high-energy variability observed in both GRBs and blazars \citep{Lazar2009, Lyubarsky2010, Giannios2009, Narayan2009}. A detailed study of this system is beyond the scope of this paper\footnote{Since the simulation lack explicit resistivity, even if 3D simulations are employed, it remains uncertain whether the numerical resistivity faithfully represents the physical resistivity \citep[but see][]{2020ApJ...900..100R}. A more robust assessment of whether such reconnection sites can efficiently dissipate magnetic energy ultimately requires fully three‑dimensional, high‑resolution, resistive GRMHD simulations.}.

Finally, we comment on the potential impact of 3D effects on the global jet dynamics. The jets in our simulations are launched into a funnel surrounded by an accretion disk and continue expanding into a low density environment, where they are largely confined by the disk wind. This configuration is valid for AGN jets and in some extent to the post-breakout phase of a GRB jet from a stellar envelope or a binary neutron star merger remnant. We show that inside the funnel the jet spine remains rather cylindrical, which is susceptible to non-axisymmetric kink modes that can generate helical distortions in the jet and affect its evolution\footnote{Jet destruction can occur also when the disk angular momentum is insufficient \citep{2025ApJ...985..135C}.} \citep[e.g.][]{Bromberg2016, Tchekhovskoy2016, 2024ApJ...964...79L}. In addition, internal kink modes can develop at the jet base in regions where collimation is strong and the jet develops a collimation nozzle \citep{Lyubarsky2009, Globus2016}. Matter passing through this nozzle may develop internal kink modes that incite reconnection within a mini-jet, dissipating part of the toroidal magnetic field energy and driving turbulence \citep{Bromberg2016, Barniol-Duran2017, Bromberg2019}. In addition, Rayleigh-Taylor (or its magnetic counterpart, Kruskal-Schwarzschild) instability can grow on the edges of the nested fingers in the cocoon or on the interface between the jet spine and the cocoon, especially in regions where the magnetic field is weak and induce turbulent mixing. The wavevector of the unstable modes is in the azimuthal direction and requires full 3D simulation to realize.

\section{Conclusions} \label{sec:conclude}

We show for the first time the dynamical evolution of a striped jet, from its launching by the accretion of magnetic loops with opposite orientation, to the formation of current sheets by the interactions between the stripes. We analyze the launching and acceleration process, lateral jet structure and the efficiency of energy dissipation in the current sheets. Our model could be appropriate for AGN jets and the post-breakout stage of GRB jets, where the jet expands into vaccum and is confined from the side by a sub relativistic outflow. Our main findings are listed below:
\begin{itemize}

\item \textit{Jet launching efficiency}: We confirm the findings of \citet{2021MNRAS.508.1241C} that efficient launching of the mini-jets require the loops to form not too far from the BH, at a distance $r < 100\,r_g$. Loops that form at larger distance will lose coherency due to turbulence before reaching the BH and launch only weak jets.

\item \textit{Mini-jets properties}: The accretion of each loop leads to the launching of a `mini-jet' that accelerates as a normal jet. The mini-jet is composed of a relativistic, PFD spine carrying energy on the outgoing fieldlines, surrounded by a sub-relativistic sheath threaded by the returning field acting as a cocoon that collimates the jet spine. The sheath is loaded with mass that seeps into the vicinity of the BH during polarity switch episodes and is pushed aside by the mini-jet once it is launched.

\item \textit{Stratified sheath structure}: Under axisymmetry, striped jets feature a nested, jet-within-a-jet configuration where each mini-jet is launched into the remnant material of the previous one. As a result, the sheath features a coaxial stratified structure, where each layer carry magnetic field with an opposite orientation to its neighbor. Since the layers are mass loaded, the dynamics is set by the plasma and therefore reconnection at the boundaries between the layers is inefficient.

\item \textit{Current sheets and energy dissipation}: Current sheets form when two mini-jets come into close contact with each other, producing a sharp reversal in the toroidal magnetic field and an inflow of magnetic energy into the interface in the sheet frame. Since the mini-jets accelerate to similar Lorentz factors, formation of current sheets sufficiently close to the BH so that the flow remain PFD requires neighboring mini-jets to catch up during the early acceleration phase, before becoming super-fast-magnetosonic. During this stage, a mini-jet expands longitudinally under its magnetic pressure, causing its tail to accelerate less efficiently than its head and allowing the head of the trailing mini-jet to gain on it. This requires formation of relatively small loops (of a size of a few $10\,r_g$) and fast polarity switch on the BH. Once the current sheet is formed it features an observed energy dissipation rate of $\mathcal{O}(0.1)$ of the BZ power during the launching period, providing enough power for the current sheet to be observed as a superluminal bright feature.

\item \textit{Current sheet lifetime}: The current sheet requires a continual supply of toroidal field to be active. Given our estimated limits on the size of a mini-jet and a typical inflow proper velocity of $v'=0.1c$, we estimate a lifetime of $\lesssim10^5\,t_g$, which is translated to $\sim10$ pc in the case of a blazar jet launched from a $10^9~M_\odot$ BH. This length scale is consistent with observations of superluminous features at $1-10$ pc scales in BL-Lac objects such as TXS 0506+056.

\end{itemize}
%


\section*{Acknowledgements}
The authors acknowledge the support of a BSF grant 2024297.
Ho-Sang (Leon) Chan acknowledges support from the Croucher Scholarship for Doctoral Studies by the Croucher Foundation. OB was supported by an ISF grant 2067/22 and an NSF-BSF grant 2024788.  GW acknowledges support from NASA grant 80NSSC24K0941.

\section*{Data Availability}
The simulation data underlying this article will be shared on reasonable request to the corresponding author.



\bibliographystyle{mnras}
\bibliography{example} 



\newpage 


\appendix

\begin{figure*}
    \centering
    \includegraphics[width=1.0\linewidth]{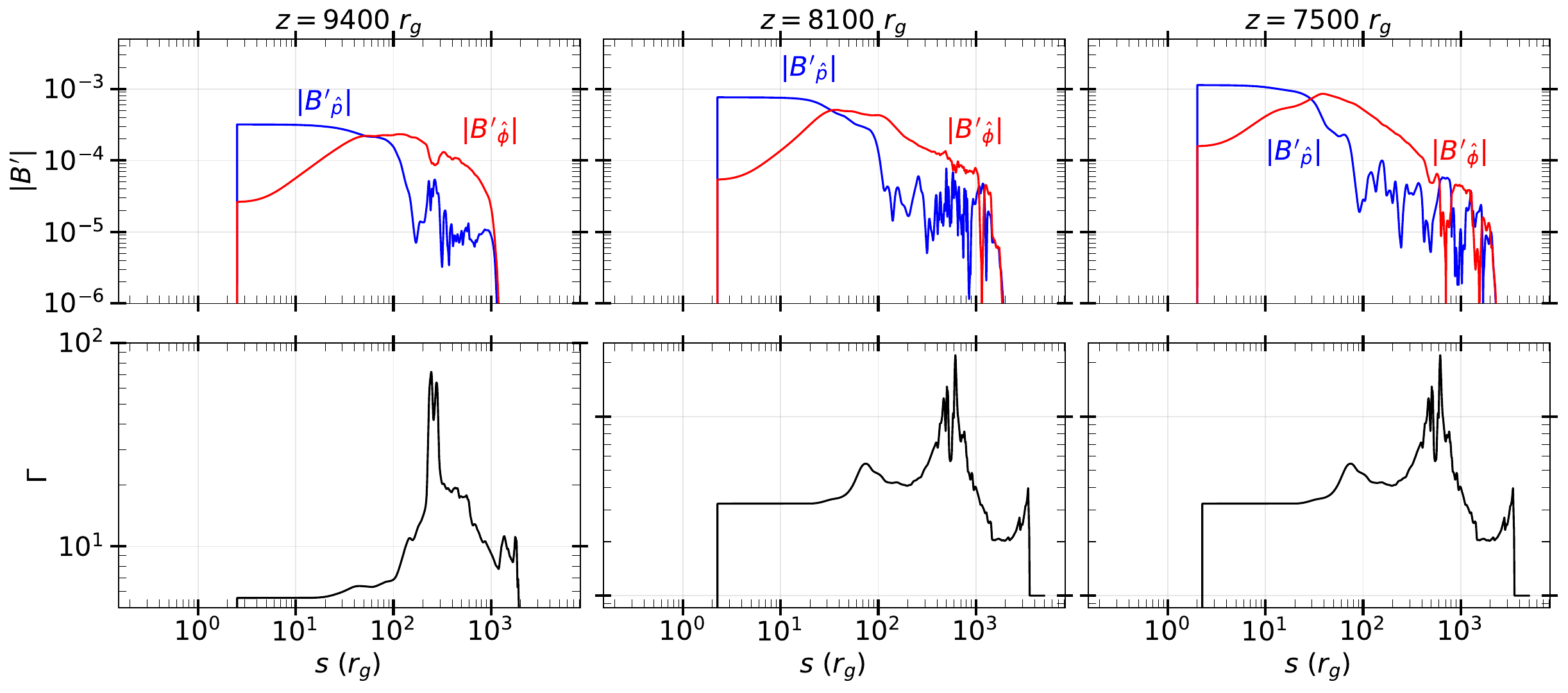}
    \caption{Upper row: Fluid-frame poloidal (blue) and toroidal (red) magnetic field strengths as functions of cylindrical radius for the $N_{\rm loop}=5$ model at $t = 10^4$\,$t_g$, measured at several heights $z$ as indicated in the column titles. Lower row: Same quantities as in the upper row, but for the gas Lorentz factor. \label{fig:appendix.pdf}}
\end{figure*}

\section{Supplementary Material} \label{sec:supp}
In Figure~\ref{fig:appendix.pdf}, we show how the fluid-frame poloidal and toroidal magnetic fields, as well as the gas Lorentz factor, vary as functions of cylindrical radius. We focus on the $N_{\rm loop}=5$ model at $t = 10^4\,t_g$ and select three representative heights $z$ at which to measure these quantities; these locations correspond approximately to the positions of the first three mini-jets. The figure indicates that, close to the $z$-axis, the poloidal field profile is flat, and it is stronger than the toroidal field. Moving outward from the axis, the poloidal field strength decreases sharply while the toroidal component increases and eventually becomes dominant, peaking roughly at the position where the poloidal field strength drops sharply.
This creates a radial range in which $|B'_{\hat{p}}| \sim |B'_{\hat{\phi}}|$. Both components decline as one exits the jet. Similarly, the gas Lorentz factor peaks off-axis with a magnitude of $\mathcal{O}(10)$. These features are broadly consistent with those of a continuous PFD jet \citep{2016MNRAS.456.1739B}.

\section{Extracting Fluid Variables} \label{sec:fluid}
It is often useful to compute vector quantities in an orthonormal basis. Relativistic jets can propagate to radii of $\gtrsim O(10^{3})$\,$r_{g}$. At such distances, the spacetime is effectively flat. Thus, we approximate the metric as Minkowski space, expressed in standard spherical coordinates. In this frame, $g_{\mu\nu} = 0$ for $\mu \neq \nu$, while $g_{00} = -1$, $g_{11} = 1$, $g_{22} = r^{2}$, and $g_{33} = r^{2}\sin^{2}\theta$. We can then trivially transform vectors from the coordinate basis to the orthonormal basis via $U^{\hat{i}} = \sqrt{g_{ii}}\,U^{i}$. Here, $\vec{U}$ denotes the $3-$vector of interest, such as the magnetic field $\vec{B}$ or the coordinate velocity $V^{i} = u^{i}/u^{t}$. Obtaining vectors in the orthonormal basis offers the advantage of allowing standard vector operations as if in flat Newtonian space. For example, the current density $\vec{J}$, electric field $\vec{E}$, and the magnitudes of the magnetic field and velocity can be written in a straightforward form ($\mu_{0} = 1$ in the cgs unit)
\begin{equation}
\begin{aligned}
    \vec{J} &= \vec{\nabla}\times\vec{B}, \\
    \vec{E} &= -\vec{V}\times\vec{B}, \\
    V^{2} &= (V^{\hat{r}})^{2} + (V^{\hat{\theta}})^{2} + (V^{\hat{\phi}})^{2}, \\
    B^{2} &= (B^{\hat{r}})^{2} + (B^{\hat{\theta}})^{2} + (B^{\hat{\phi}})^{2}.
\end{aligned}
\end{equation}
The above considerations hold if the coordinate system is Boyer-Lindquist. However, \texttt{H-AMR} evolves the system using a horizon-penetrating Kerr-Schild coordinate system. In this coordinate frame, there exists a non-zero off-diagonal component, $g_{13}$, which persists even in the asymptotic regime $r \gg 1$ --- unless the BH is non-spinning. This term arises due to an infinite untwisting of the temporal-azimuthal coordinate, designed to maintain regularity at the event horizon. To eliminate this term, we transform four-vectors from Kerr-Schild to Boyer-Lindquist coordinates via
\begin{equation}
\begin{aligned}
    u_{\rm BL}^{0} &= u_{\rm KS}^{0} - \frac{2r}{r^{2} - 2r + a^{2}}u_{\rm KS}^{1}, \\
    u_{\rm BL}^{1} &= u_{\rm KS}^{1}, \\
    u_{\rm BL}^{2} &= u_{\rm KS}^{2}, \\
    u_{\rm BL}^{3} &= u_{\rm KS}^{3} - \frac{a}{r^{2} - 2r + a^{2}}u_{\rm KS}^{1},
\end{aligned}
\end{equation}
where $u_{\rm BL}^{\mu}$ ($u_{\rm KS}^{\mu}$) denotes the $4-$velocity in Boyer-Lindquist (Kerr-Schild) coordinates.

For the magnetic field, we similarly transform the $4-$field $b^{\mu}$ from the Kerr-Schild to the Boyer-Lindquist coordinates, and then extract the $3-$vector field $B^{i}$ as
\begin{equation}
    B^{i} = b_{\rm BL}^{i}u_{\rm BL}^{t} - b_{\rm BL}^{t}u_{\rm BL}^{i}, 
\end{equation}
so that once both $B^{i}$ and $V^{i}$ are obtained, we can proceed with the operations outlined above.

In some cases, we want to evaluate the vector quantities in the rest frame of the fluid along the striped jets, which move relativistically. To accomplish this, we perform the flat-space relativistic transformation of $\vec{J}$, $\vec{B}$, and velocity $\vec{U}$ as follows
\begin{equation} \label{eqn:transform}
\begin{aligned}
    \vec{B'} &= \Gamma\left(\vec{B} - \vec{V}\times\vec{E}\right) - (\Gamma - 1)\frac{(\vec{B}\cdot\vec{V})\vec{V}}{|V|^{2}}, \\
    \vec{J}' &= \vec{J} + (\Gamma - 1)\frac{(\vec{J}\cdot\vec{V})\vec{V}}{|V|^{2}}, \\
    \vec{U}' &= \frac{1}{1 - \vec{V}\cdot\vec{U}}\left[\frac{\vec{U}}{\Gamma} - \vec{V} + \frac{\Gamma}{\Gamma + 1}(\vec{U}\cdot\vec{V})\vec{V}\right],
\end{aligned}
\end{equation}
where $\vec{V}$ is the frame that one would like to boost into and $\Gamma$ here is the corresponding Lorentz factor. Note that the vector quantities in the above equation should first be transformed from the coordinate basis to the orthonormal basis. This approximation should be valid for $r \geq 100$\,$r_g$ where $M/r \leq 0.01$.

\section{Calculating the observed power of the current sheet}
\label{sec:CS_Power}
Given an energy dissipation rate $\dot{E'}$ in the frame of the current sheet and assuming that a fraction $\eta$ is converted into radiation, we would like to calculate the observed isotropic equivalent luminosity and one sided jet power seeing by an on axis observer. We assume a conical jet with an opening angle $\theta_j\ll1$ carrying a relativistically moving current sheet with Lorentz factor $\Gamma$ that radiates uniformly and isotropically in its proper frame. The bolometric intensity in the frame of the current sheet is \citep{Rybicki1979}
\begin{equation}\label{App:I}
I'=\frac{dP'}{dA_\perp d\Omega'}=\eta\frac{\dot{E}'}{\pi r_{_{\rm CS}}^2\Omega_j}\sim\eta\frac{\dot{E}'}{\pi^2r_{_{\rm CS}}^2\theta_j^2},
\end{equation}
where $P'$ is the power, $\Omega_j=2\pi(1-\cos\theta_j)\simeq\pi\theta_j^2$ is the jet solid angle and $r_{_{\rm CS}}$ is the radial distance of the current sheet from the BH.
Conversion to the observer frame gives
\begin{equation}
I=\delta^4 I'\simeq\Gamma^4I'.
\end{equation}
Due to relativistic beaming the observer can only see light coming from a solid angle $\Omega_b\simeq\pi\theta_b^2$, where $\theta_b=\Gamma^{-1}$ is the beaming angle, thus the observed flux can be estimated as 
\begin{equation}\label{App:Fobs}
F_{\rm obs}=\frac{r_{_{\rm CS}}^2}{D^2}\int_{\Omega_b}I\cos\theta d\Omega
\simeq \frac{r_{_{\rm CS}}^2}{D^2}\frac{\pi I}{\Gamma^2},
\end{equation}
where $D$ is the distance to the observer, neglecting cosmological expansion. Combining Equation~\ref{App:I}-\ref{App:Fobs}, the associated isotropic equivalent luminosity and angle corrected jet power are
\begin{equation}
L_{\rm iso}=4\pi D^2 F_{\rm obs}=4\Gamma^2\frac{\eta\dot{E}'}{\theta_j^2},
\end{equation}
and
\begin{equation}\label{App:Lj}
L_{\rm j}=\Omega_j D^2 F_{\rm obs}=\Gamma^2\eta\dot{E}',
\end{equation}
respectively.


\bsp	
\label{lastpage}
\end{document}